\documentclass[journal,onecolumn,draft,11pt]{IEEEtran}

\usepackage[dvips]{graphicx}
\usepackage{epsfig}
\usepackage[cmex10]{amsmath}
\usepackage{amssymb}
\usepackage{amsthm}
\usepackage{amsfonts}
\usepackage{bm}

\usepackage{cite}
\usepackage[svgnames]{xcolor}
\usepackage[tight,footnotesize]{subfigure}
\usepackage{algpseudocode}

\usepackage{algorithm}

\graphicspath{{figs/}}

\newcommand{\defeq}{\mathrel{\triangleq}}

\newcommand{\Cb}{\mathbf{C}}
\newcommand{\Ab}{\mathbf{A}}
\newcommand{\Bb}{\mathbf{B}}

\newcommand{\iid}{i.\@i.\@d.\ }

\newtheorem{lemma}{Lemma}

\newtheorem{theorem}[lemma]{Theorem}

\theoremstyle{definition}

\newtheorem{egdummy}{Example}

\newtheoremstyle{myremark}
{\topsep}{\topsep}{\normalfont}{\parindent}{\itshape}{:}{ }{}

\theoremstyle{myremark}

\newcounter{Remark}
\newenvironment{Remark}
{
	\refstepcounter{Remark}
	\textbf{Remark \theRemark:}
}

\newcounter{Algorithm}
\newenvironment{Algorithm} [1]
{
	\refstepcounter{Algorithm}
	\hrule
	\textbf{Algorithm \theAlgorithm: #1 }
	\hrule
}
{
	\hrule}

\newcommand\shortintertext[1]{
	\ifvmode\else\\\@empty\fi
	\noalign{
		\penalty0
		\vbox{\mathstrut}
		\penalty10000
		\vskip-\baselineskip
		\penalty10000
		\vbox to 0pt{
			\normalbaselines
			\ifdim\linewidth=\columnwidth
			\else
			\parshape\@ne
			\@totalleftmargin\linewidth
			\fi
			\vss
			\noindent#1\par}
		\penalty10000
		\vskip-\baselineskip}
	\penalty10000}

\makeatletter
\def\endthebibliography{
	\def\@noitemerr{\@latex@warning{Empty `thebibliography' environment}}
	\endlist
}
\makeatother

\begin{document}
	
\title{Coded Secure Multi-Party Computation for Massive Matrices with Adversarial Nodes}

\author{Seyed Reza Hoseini Najarkolaei$^{*}$, Mohammad~Ali~Maddah-Ali$^{\dagger}$, Mohammad Reza Aref$^{*}$
	\\{$^{*}$ Department of Electrical Engineering, Sharif University of Technology, Tehran, Iran}
	\\{$^{\dagger}$ Nokia Bell Labs, Holmdel, NJ, USA}
	
}

\maketitle
\begin{abstract} 
In this work, we consider the problem of secure multi-party computation (MPC), consisting of $\Gamma$ sources, each has access to a large private matrix,  $N$ processing nodes or workers, and one data collector or master. The master is interested in  the result of a polynomial function of the input matrices.   Each source sends a randomized functions of its matrix, called as its share, to each worker. The workers process their shares in interaction with each other, and send some results to the master such that it can derive the final result. There are several constraints: (1) each  worker can store a function of each input matrix, with the size of $\frac{1}{m}$ fraction of that input matrix, (2) up to $t$ of the workers, for some integer $t$, are adversary and may collude to gain information about the private inputs or can do malicious actions to make the final result incorrect. 
The objective is to design an MPC scheme with the minimum number the workers, called the recovery threshold, such that the final result is correct, workers learn no information about the input matrices, and the master learns nothing beyond the final result.  In this paper, we propose an MPC scheme that achieves the recovery threshold of $3t+2m-1$ workers, which is order-wise less than the recovery threshold of  the conventional methods. The challenge in dealing with this set up is that  when nodes interact with each other, the malicious messages that adversarial nodes generate propagate through the system, and can mislead the honest nodes. To deal with this challenge, we design some subroutines that can detect erroneous messages, and correct or drop them.

\end{abstract}
\begin{IEEEkeywords}
    multi-party computation, polynomial codes, secure computation, massive matrix computation.
\end{IEEEkeywords}
\section{Introduction}
\label{sec:introduction}

In many applications, preserving privacy of  data is one of the major concerns, when the computation is outsources to some external nodes.  
 
Secure bidding and auction\cite{bogetoft2006practical}, secure E-voting\cite{nair2015improved}, preserving privacy data mining\cite{narwaria2016privacy}, sharing of signature \cite{franklin1995verifiable}, and private information retrieval \cite{chor1995private} are some applications that preserving privacy of data is indeed the main concern. 
Processing private inputs on several nodes where some of them are untrusted is studied in the context of secure multi-party computation. Multi-party computation is introduced to be able to distribute processing among some nodes, without leaking any information, even though some of them may collude or behave adversarial~\cite{ben2008fairplaymp,galil1987cryptographic,canetti1996adaptively,goldreich1998secure,gennaro1998simplified,ben1988completeness,nodehi2018entangled,mypaper,nodehi2019secure,pedersen1991non,rabin1989verifiable,cramer2000general}. Security of a multi-party computation systems can be guaranteed cryptographically~\cite{gennaro1998simplified,galil1987cryptographic,canetti1996adaptively,ben2008fairplaymp},  or  information theoretically \cite{ben1988completeness, nodehi2018entangled,mypaper,nodehi2019secure,pedersen1991non,rabin1989verifiable,goldreich1998secure}. In a cryptographically secure systems, breaking the encryption system requires infeasible computation power, and is not practically possible. 
In an information theoretically secure system, adversarial nodes can not gain any information about the secret, even if they have unlimited computing power. In this paper, the objective is to design an information-theoretically secure MPC scheme, for the cases where the size of the input data is massive, and thus the computation has to be distributed among some nodes, where some of them are curious about the input, or adversary seeking to make the final result incorrect. 

Recently, coding techniques has been developed to deal with another important challenge in distributing massive computing to external nodes, which is latency of the system due stragglers~\cite{dean2013tail, lee2018speeding, lee2017high, yu2017polynomial,yu2018straggler,dutta2018optimal,jahani2018codedsketch}. 
In distributed systems, to have the final result of computation, we need to  wait for slowest nodes, or \emph{stragglers}~\cite{dean2013tail}.

To solve this problem, in \cite{lee2018speeding} and \cite{lee2017high}, it is proposed to use Maximum Distance Separable (MDS) codes to inject coded redundancy in computing, and thus avoid waiting for the slowest nodes.  
In \cite{yu2017polynomial},  \emph{polynomial codes} is developed for distributed massive matrix multiplications, where the results of computations at the processing nodes form an optimum code,  minimizing the number of workers that we need to wait for, known as the \emph{recovery threshold}. Polynomial codes are extended to \emph{entangled polynomial codes} and \emph{Matdot code} in \cite{nodehi2018entangled,dutta2018optimal} to admit more flexible partitioning of each input matrix. 

In \cite{jahani2018codedsketch},  \emph{CodedSketch} is proposed which also enables us to compute an approximation of the multiplication of two massive matrices, and reduces the recovery threshold depending on the required accuracy. 
In addition, in \cite{yu2018lagrange},   \emph{Lagrange codes} are proposed, using a different approach,  to embed coded redundancy in computing parallel computations of one arbitrary polynomial function for different inputs, where computation of the the function for one input is within computation and storage capacity of one worker.

Ideas from coded computing have been used to develop MPC for massive data~\cite{mypaper, nodehi2018entangled,nodehi2019secure,d2020gasp,yang2018secure,kim2019private,aliasgari2019private,yu2020entangled,chang2018capacity,chang2019upload,jia2019capacity,ebadifar2019need, yu2018lagrange,chen2020gcsa}. 
 In \cite{mypaper,nodehi2018entangled,nodehi2019secure}, a scheme for MPC is proposed to compute arbitrary polynomial functions of massive matrices. The scheme of \cite{mypaper,nodehi2018entangled,nodehi2019secure} is based a new sharing scheme, called polynomial sharing, which is inspired by polynomial coding~\cite{yu2017polynomial}. The number of workers needed for this computation is order-wise less than conventional approaches which are based on job-splitting.  The scheme of \cite{mypaper,nodehi2018entangled,nodehi2019secure} is developed for the cases, where some of the nodes are semi-honest, meaning that they follow the protocol, but may collude to gain information about the inputs. 
Another closely related, and parallel,  line of work is known as secure matrix multiplication~\cite{chang2019upload,yu2020entangled,kim2019private,yang2018secure,aliasgari2019private,chang2018capacity,jia2019capacity,ebadifar2019need,d2020gasp}, where the focus is to compute the result of matrix multiplication without leaking any information to the workers. 
In addition,  Lagrange codes \cite{yu2018lagrange} can handle private computation of many parallel computation of the same function for different inputs, for the cases where some workers are colluding or adversarial. Unlike \cite{mypaper,nodehi2018entangled,nodehi2019secure}, in both secure matrix multiplication~\cite{chang2019upload,yu2020entangled,kim2019private,yang2018secure,aliasgari2019private,chang2018capacity,jia2019capacity,ebadifar2019need,d2020gasp} and Lagrange codes \cite{yu2018lagrange}, there is no \emph{interaction} among the workers. The advantage is that the data collector can deal with  \emph{adversarial nodes}, at the decoding level, at the cost of increasing the recovery threshold. The disadvantage is that for secure matrix multiplication, the scheme is limited to pairwise matrix multiplication, and for Lagrange codes \cite{yu2018lagrange}, the recovery threshold grows linearly with the degree of the polynomial.  
The main objective of this paper is to develop an MPC for an arbitrary polynomial function, for the cases where some of the workers are adversarial. In this formulation, like \cite{mypaper,nodehi2018entangled,nodehi2019secure}, the workers interact with each other, which makes managing adversarial nodes challenging. Later, in Section \ref{sec:Problem Setting}, we will explain the basic differences of this work with secure matrix multiplication and Lagrange codes. 

In this paper, we consider an MPC system, including some sources, where each of them has access to a large and private  matrix. There is a data collector that is interested in an arbitrary polynomial function of the input matrices. For computing resources, there are some workers that can interact with each other.  Each worker can take a function of  each input, with the size of $\frac{1}{m}$ fraction of each input matrix. All workers are honest, except up to $t$ of them, for some integer $t$, who are malicious. This means that they can collude to gain information about the input and/or do whatever they can to make the final result incorrect. This includes sending misleading information to other workers. The objective is to design an MPC scheme, such that the data collector can recover the function correctly, in presence of  adversarial behavior of the malicious nodes. In addition, the workers gain no information about the input data. Moreover the data collector gain no information about the input matrices, beyond the result of function computation. We propose scheme that achieves the recovery threshold of $3t+2m-1$ workers, which is order-wise less than the recovery threshold of  the conventional methods.

The rest of the paper is organized as follows. In~Section \ref{sec:Problem Setting} we state the problem formulation. 
In Section \ref{main result}, we state the main result. 
In Section \ref{polynomialsharing} we define polynomial sharing.
In Section \ref{without}, we explain a new method for multiplication of two matrices, which works for semi-honest setting. To make this scheme work in the presence of  malicious  workers, we need some preliminaries and fault-tolerant subroutines, including extended verifiable secret sharing  explained in Section \ref{preliminaries}. Then, in Section \ref{with}, Section  \ref{transformingshares} and Section \ref{computinganarbitrarypolynomial}, we explain the proposed MPC scheme in the presence of  malicious  workers, and prove its correctness. 
 Finally we prove the privacy of the proposed algorithm in Section \ref{security}.

\section{Problem Setting}
\label{sec:Problem Setting}
We consider a system, including $\Gamma$ sources, $N$ processing nodes or workers and one data collector or master. There is not any communication between sources but each source is connected to each of the workers. Each source $\gamma \in \{1,2,...,\Gamma\}$ has access to a matrix  $\mathbf{X}^{[\gamma]}$, where $\mathbf{X}^{[\gamma]}$ is chosen independently and uniformly at random from $\mathbb{F}^{z_{\gamma} \times v_{\gamma}}$, for some integers ${z_{\gamma} \text{ and } v_{\gamma}}$ and finite field $\mathbb{F}$.
We assume each worker has access to a function of $\mathbf{X}^{[\gamma]}$, with the size of $\frac{1}{m_{\gamma}}$ fraction of it size, for some integer $m_{\gamma}$.
There is a point to point private link between each pair of workers. Also there is an authenticated broadcast channel among all workers such that the identity of the broadcaster is known. All the workers are connected to the master, which is interested in computing an arbitrary polynomial function of $\mathbf{X}^{[1]},\mathbf{X}^{[2]},...,\mathbf{X}^{[\Gamma]}$ denoted by $\mathbf{Y}\defeq\mathbf{G}(\mathbf{X}^{[1]},\mathbf{X}^{[2]},...,\mathbf{X}^{[\Gamma]})$. 

We assume that an arbitrary subset $\mathcal{S}=\{s_1,s_2,\dots,s_t\}$ of workers are malicious, for some integer $t\leq N$. It means that they can collude and share their data with each other. In addition malicious workers may not follow the protocol. Indeed, they may do anything,  including misleading the other workers, to make the computation result incorrect.

Assume that all existing links are error-free and secure. In this system we follow a three-step solution as follows:

\begin{enumerate}
	\item \textbf{Sharing}: In this step, source $\gamma$ sends $\mathbf{\widetilde{X}}_{\gamma,j} = \mathbf{F}_{\gamma,j}(\mathbf{X}^{[\gamma]})$ to the worker $j$, where $\mathbf{F}_{\gamma,j}(\mathbf{X}): \mathbb{F}^{z_{\gamma} \times u_{\gamma}} \rightarrow \mathbb{F}^{p \times q}$, for some integer parameters $p \text{ and }q$ and $j \in \{1,2,...,N\}$. We defined $\frac{1}{m_{\gamma}} \triangleq \frac{ pq }{ z_{\gamma} \times u_{\gamma} }$ as \emph{normalized storage capacity} of source $\gamma$ to each worker. Thus,  $\frac{1}{m_{\gamma}}$ is a real non-negative number between 0 and 1.
	
	\item \textbf{Computation and Communication}: Workers process their input data and communicate with each other in this step. Let $\mathcal{M}_{ n \rightarrow n'}$ be the set of all messages that worker $n'$ receives from worker $n$ in this step, for some $n,n' \in \{1,2,...,N\}$. In the main protocol, for ease of understanding, computation and communication step will be partitioned into 2 phases, subsharing phase, and constructing the product of subshares phase.
	
	\item \textbf{Reconstruction}: In this step, each worker $n$,  $n \in \{1,2,...,N\}$, sends a message, denoted by $\mathbf{Y}_n$ to the master, such that the master is able to reconstruct $\mathbf{Y}=\mathbf{G}(\mathbf{X}^{[1]},\mathbf{X}^{[2]},...,\mathbf{X}^{[\Gamma]})$ from $\mathbf{Y}_1,\mathbf{Y}_2, ... \mathbf{Y}_{N}$.
\end{enumerate}

All schemes that are proposed in secure fault-tolerant MPC systems must satisfy some constraints regarding, correctness, privacy for the workers, and privacy for the master, as follows.

\begin{itemize}
	\space \item	\textbf{Correctness}: Master must have sufficient information to derive the correct result. More precisely,
	\begin{align}
	H(\mathbf{Y}|\mathbf{Y}_1,\mathbf{Y}_2,...,\mathbf{Y}_{N})=0.
	\end{align}
 Note that the correctness condition must be satisfied in the presence of adversarial nodes. 
\end{itemize}

\begin{itemize}
\label{secondC}
	\space \item	\textbf{Privacy for the workers}: Let $ t \in \{1,...,N\}$. If any arbitrary subset $\mathcal{S}=\{s_1,s_2,...,s_{t}\}$ of workers collude and share their data, they can not gain any information about the inputs, i.e., 
    \begin{align*}
    \label{Privacyfortheworkers}
            H(\mathbf{X}^{[j]},j \in \{1,2,...,\Gamma\})=
        H(\mathbf{X}^{[j]},j \in \{1,2,...,\Gamma\}|
        \displaystyle\cup_{s_i \in \mathcal{S}}\{&\mathcal{M}_{n\rightarrow s_i}, n \in\{1,2,...,N\}\},\mathbf{\widetilde{X}}_{\gamma,s_i},\\&
        \gamma\in\{1,2,...,\Gamma\},s_i \in \mathcal{S})
    \end{align*}
\end{itemize}
 
\begin{itemize}
\label{thirdC}
	\space \item	\textbf{Privacy for the master}: Master must not gain any additional information beyond the function result, i.e.,
	\begin{align*}
	    H(\mathbf{X}^{[1]},\mathbf{X}^{[2]},...,\mathbf{X}^{[\Gamma]}|\mathbf{Y},\mathbf{Y}_1,\mathbf{Y}_2,...,\mathbf{Y}_{N})=
	    H(\mathbf{X}^{[1]},\mathbf{X}^{[2]},...,\mathbf{X}^{[\Gamma]}|\mathbf{Y}).
	\end{align*}
\end{itemize}

\begin{figure}[htbp]
    \centering
    \includegraphics[draft=false,width=85mm]{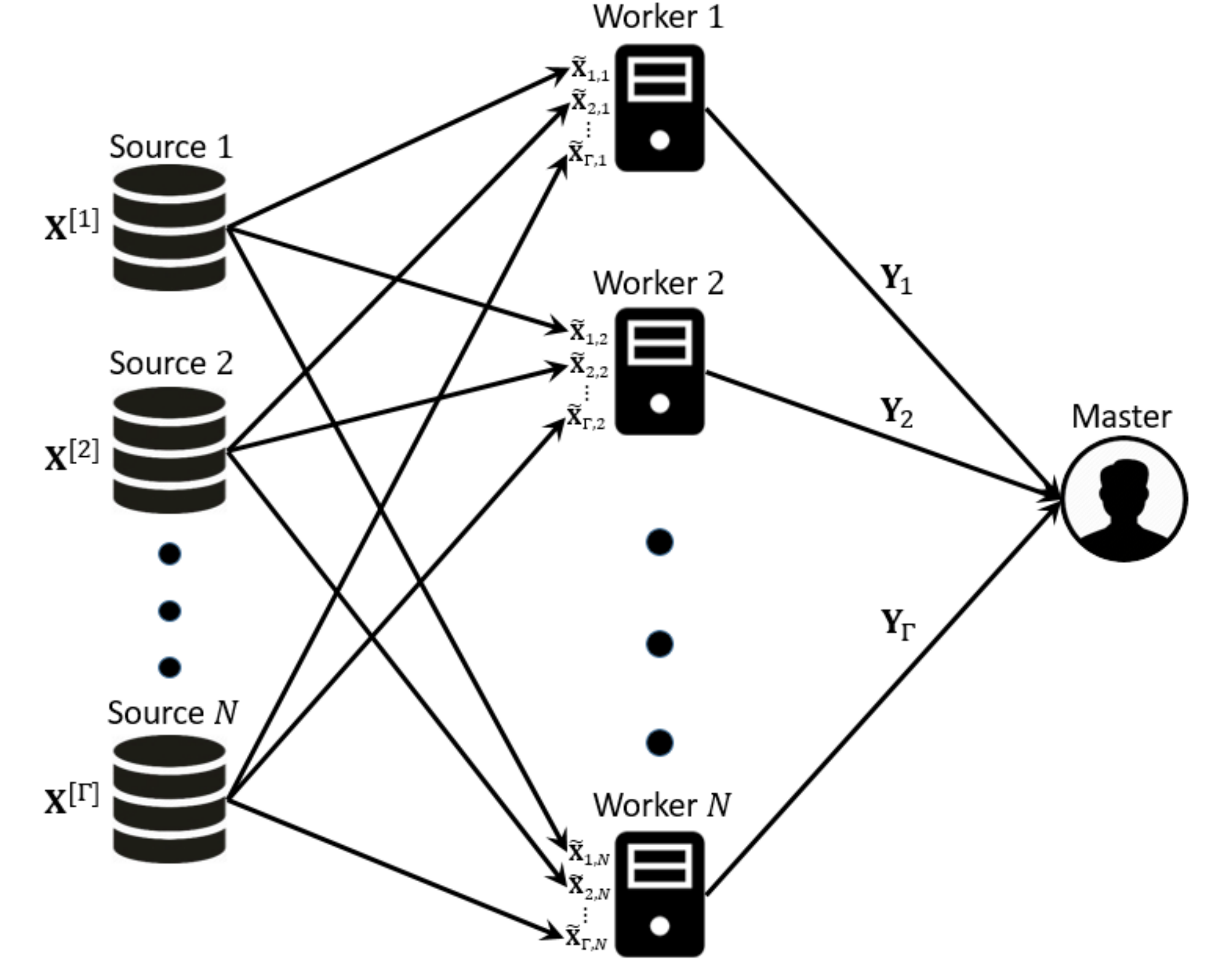}
    \caption{ An MPC system includes $\Gamma$ sources, $N$ processing nodes or workers, and one data collector or master. There is not any communication between sources but each source is connected to each of the workers. Each source $\gamma \in \{1,2,...,\Gamma\}$ has access to a large-scale matrix  $\mathbf{X}^{[\gamma]}$. We assume that each worker can store a function of $\mathbf{X}^{[\gamma]}$, with the size of $\frac{1}{m_{\gamma}}$ fraction of it size for some integer $m_{\gamma}$. There is a point to point private link between each pair of workers. Also there is an authenticated broadcast channel among all workers such that the identity of the broadcaster is known. All the workers are connected to the master, which is interested in computing an arbitrary polynomial function of $\mathbf{X}^{[1]},\mathbf{X}^{[2]},...,\mathbf{X}^{[\Gamma]}$ denoted by $\mathbf{Y}\defeq\mathbf{G}(\mathbf{X}^{[1]},\mathbf{X}^{[2]},...,\mathbf{X}^{[\Gamma]})$.}
    \label{fig1}
\end{figure}

\newtheorem{mydef}{\textbf{Definition}}

\begin{mydef}
   Consider the secure MPC system, explained above, with $t$ malicious workers and $m_\gamma=m$,  for $\gamma \in [\Gamma]$, and some integer $m$. The \emph{recovery threshold}, denoted by $N^*_{t,\frac{1}{m}}$,  is defined as the minimum number of workers $N$, needed to calculate an arbitrary polynomial function of private inputs $\mathbf{Y}=\mathbf{G}(\mathbf{X}^{[1]},\mathbf{X}^{[2]},...,\mathbf{X}^{[\Gamma]})$, while satisfying the correctness condition,  the privacy condition at the workers,  and  the privacy condition at the data collector.
   \end{mydef}
\begin{Remark}
In this paper, the objective is to introduce an upper-bound on $N^*_{t,\frac{1}{m}}$, by developing an achievable scheme for this problem. 
\end{Remark}

\begin{Remark}
 The difference between this set-up and  secure matrix multiplication \cite{chang2019upload,yu2020entangled,kim2019private,yang2018secure,aliasgari2019private,chang2018capacity,jia2019capacity,ebadifar2019need,d2020gasp} is that in secure matrix multiplication, the master is interested in the result of multiplication of two matrices, or multiple pairs of matrices,  and there is no privacy constraint at the master. Also there is no communication among the workers. Thus adversarial behavior of malicious workers can be detected and corrected using schemes  similar as what is used in Reed-Solomon decoding. However, in this paper,  workers communicate with each other to be able to calculate any polynomial function, rather than just matrix multiplication.  The challenge is then to prevent adversarial nodes to mislead other workers, and also prevent errors  propagating in the network. 
\end{Remark}

\begin{Remark}
 The difference between this set-up and  Lagrange coding \cite{yu2018lagrange} is that Lagrange code is  design to compute $\mathbf{G}(\mathbf{X}_1), \ldots, \mathbf{G}(\mathbf{X}_{\Gamma})$, for some polynomial function $\mathbf{G}$ and some input matrices $\mathbf{X}_1, \ldots,  \mathbf{X}_{\Gamma}$. In addition, each worker can handle computing  one of these tasks. Lagrange code introduces coded redundancy in executing those tasks.  The algorithm can also protect privacy of data at the workers.  Lagrange codes  is a single-shot scheme, without any interaction among the workers. The advantage  thus is that adversarial nodes can be handled at the master, following the same approaches used in Reed-Solomon decoding.
The disadvantage is that the recovery threshold grows linearly with the degree of $\mathbf{G}$. In contrary, in the proposed problem formulation here, workers can communicate with each other, and thus the recovery threshold does not increase with the degree of polynomial. This makes dealing with adversarial nodes more challenging. 
\end{Remark}

\section{main result}\label{main result}

The objective is to calculate an arbitrary polynomial function $\mathbf{G}$ of the inputs. As explained in \cite{nodehi2018entangled}, any arbitrary polynomial function can be implemented using a sequence of multiplication, addition, and transposing gates. In this paper a new scheme called polynomial coded multi-party computation is proposed, which order-wise reduces the recovery threshold as compared with conventional schemes. In Section \ref{with}, we see how to compute the product of two matrices while the privacy and correctness conditions have been satisfied.
We state the main result in the following theorem.
	\begin{theorem}\label{remark}
		Consider matrices $\mathbf{X}^{[1]},\mathbf{X}^{[2]},...,\mathbf{X}^{[\Gamma]}$ where $\mathbf{X}^{[\gamma]}$ are chosen independently and uniformly at random from $\mathbb{F}^{k \times k}$, for some integer $k$. Assume each worker can take a function of  each input, with the size of $\frac{1}{m}$ fraction of that input matrix.
		The optimum recovery threshold to compute an arbitrary polynomial function $\mathbf{G}(\mathbf{X}^{[1]},\mathbf{X}^{[2]},...,\mathbf{X}^{[\Gamma]})$ is upper-bounded by  $2m+3t-1$.
	\end{theorem}		
Proof can be found in Section \ref{with}, which is based on the procedures, provided in the rest of the paper.

\begin{Remark}
	To the state of the art, the proposed scheme order-wise reduces the recovery threshold as compared to conventional methods \cite{ben1988completeness}. 		
	For example, consider a secure MPC system with $t=m=20$. Concatenating job splitting and conventional BGW scheme needs more than $1200$ workers, while the proposed scheme needs just $99$ workers.
\end{Remark}

\begin{Remark}
	As it can be seen, the upper bound of the recovery threshold is a linear function of $m$ and $t$. The fact that the upper bound is a function of 3 times the number of adversarial nodes is a common phenomenon in  distributed systems with adversarial nodes.	
\end{Remark}

\begin{Remark}
	Unlike secure matrix multiplication methods~\cite{chang2019upload,yu2020entangled,kim2019private,yang2018secure,aliasgari2019private,chang2018capacity,jia2019capacity,ebadifar2019need,d2020gasp}, the proposed scheme is not limited to pair-wise multiplication of matrices. In addition, unlike Lagrange coded computing \cite{yu2018lagrange}, the recovery threshold does not increase with the degree of the polynomial. That is because in this setting, workers can communicate each other, which raises new challenges, mentioned in the next remark.
\end{Remark}

\begin{Remark}
	 The challenge that we face in this setting is that the errors of adversaries can propagate through the entire system and even mislead the honest nodes and make the final result incorrect. In addition, in interaction among the workers, there is the risk of leaking information about private inputs to the workers. Therefore, we need to develop some subroutines that protect us against colluding or malicious actions. For this reason, we introduce a new method of computation in Section \ref{without} that can be combined with some protecting subroutines, explained in Section \ref{preliminaries}, and resolve these challenges. For the case where $m=1$, this scheme reduces to BGW protocol \cite{ben1988completeness}.
\end{Remark}

\section{Polynomial Sharing}
\label{polynomialsharing}
\begin{mydef}
\label{definition}
Consider matrix $\mathbf{X} \in \mathbb{F}^{z \times v}$, for some $z,v \in \mathbb{N}$, partitioned as 
\begin{align}
\mathbf{X}=\begin{bmatrix}
\mathbf{X}_{0},\mathbf{X}_{ 1},\mathbf{X}_{ 2},\dots,\mathbf{X}_{m-1}
\end{bmatrix},
\end{align}

where $\mathbf{X}_{i} \in \mathbb{F}^{z \times \frac{v}{m}}$, $i=1, \ldots, m-1$,  for some $m \in \mathbb{N}$ where  $m|v$.
We define direct polynomial coded matrix $\mathbf{F}^{(\mathsf{d})}_{\mathbf{X},m,t,\delta}(x)$  and reverse polynomial coded matrix $\mathbf{F}^{(\mathsf{r})}_{\mathbf{B},m,t,\delta}(x)$ as
\begin{align*}
   &\mathbf{F}^{(\mathsf{d})}_{\mathbf{X},m,t, \delta}(x) \defeq
    \displaystyle\sum_{j=0}^{m-1} \mathbf{X}_{j}x^{j} +\sum_{q=0}^{t-1}\mathbf{R}_q x^{m+\delta+q},\\
    &\mathbf{F}^{(\mathsf{r})}_{\mathbf{X},m,t,\delta}(x) \defeq
    \displaystyle\sum_{j=0}^{m-1} \mathbf{X}_{m-1-j}x^{j} +\sum_{q=0}^{t-1}\hat{\mathbf{R}}_q x^{m+\delta+q},
\end{align*}
for some $\delta \in \mathbb{N} \cup \{0 \}$ and $q \in \{0,1,...,t-1\}$, where $\mathbf{R}_q$  and $\hat{\mathbf{R}}_q$ are chosen independently and uniformly at random from $\mathbb{F}^{z \times \frac{v}{m}}$.

Assume that each worker is assigned a distinct nonzero $\alpha_n \in \mathbb{F}$, $n  \in [N]$. We say that $\mathbf{X}$ is shared in the form of $(m,t,\delta,\mathsf{d})$-share with worker $n$,  if $\mathbf{F}^{(\mathsf{d})}_{\mathbf{X},m,t,\delta}(\alpha_n)$ is sent to worker $n$. Similarly, we say $\mathbf{X}$ is shared in the form of $(m,t,\delta,\mathsf{r})$-share with worker $n$,  if $\mathbf{F}^{(\mathsf{r})}_{\mathbf{X},m,t,\delta}(\alpha_n)$ is sent to worker $n$. We call  $\mathbf{F}^{(\mathsf{d})}_{\mathbf{X},m,t,\delta}(\alpha_n)$ and $\mathbf{F}^{(\mathsf{r})}_{\mathbf{X},m,t,\delta}(\alpha_n)$ as shares of matrix $\mathbf{X}$. 
\end{mydef}

Here we have three properties:  (1) It is shown that if $\mathbf{X}$ was shared in the form of $(m,t,\delta,\mathsf{d})$-share or $(m,t,\delta,\mathsf{r})$-share with some workers, then if less than or equal to $t$ workers collude, they gain no information about $\mathbf{X}$ \cite{nodehi2019secure}. On the other hand, with the shares of $t+m$ or more workers, we can recover matrix $\mathbf{X}$. (2) the size of $\mathbf{F}^{(\mathsf{d})}_{\mathbf{X},m,t,\delta}(\alpha_n)$ and $\mathbf{F}^{(\mathsf{r})}_{\mathbf{X},m,t,\delta}(\alpha_n)$ are $\frac{1}{m}$ fraction of the size of $\mathbf{X}$, (3) this form of sharing admits basic operations such as matrix multiplication and addition as we will show in the next sections. 

\section{Multiplication of two matrices in semi-honest setting}
\label{without}
As discussed in Section \ref{sec:introduction}, we are interested in computing an arbitrary polynomial of input matrices, under some privacy constraints.    In this section, we first focus on matrix multiplication, due to its major role in the proposed scheme. However, for now we assume that up to $t$ workers are \emph{semi-honest}, rather than adversary.   Semi-honest nodes follow the protocol, but may collude to gain some information about the inputs. This case has been addressed before in~\cite{mypaper, nodehi2018entangled,nodehi2019secure}. Here, we propose a different solution, that can be modified to handle adversarial nodes, as it will be explained later in Section \ref{with}. 

Assume that we have $\Gamma=2$ input matrices,  $\mathbf{X}^{[1]} = \mathbf{A}$ and $\mathbf{X}^{[2]}=\mathbf{B}$. At the first step  $(m,t,0, \mathsf{d})$-share of $\mathbf{A}$ and  $(m,t,0, \mathsf{r})$-share of $\mathbf{B}$ are sent to the workers. Finally, we want   $(m,t,0,\mathsf{d})$-share of $\mathbf{C}\defeq\mathbf{A}\mathbf{B}^T$, or alternatively  $(m,t,0,\mathsf{d})$-share of $\mathbf{C}$ be available, at the workers. Having shares of $\mathbf{C}$ available at the workers has two advantages: (1) These shares can be sent to the master, and it can recover file $\mathbf{C}$, while privacy at the master is held. (2) If  our goal is to multiply $\mathbf{C}=  \mathbf{A}\mathbf{B}^T$ to another matrix, we will use the same scheme again. Thus, we can calculate multiplication of any number of input matrices, and even any polynomial of them, by iteratively using the same algorithm. 
This will be discussed later in details. 

For simplicity, assume that $\Ab,\Bb \in \mathbb{F}^{z \times z}$.
The steps of the algorithm is as follows: 

 \textbf{1- Sharing phase:} Source 1 shares $\mathbf{A}$ in the form of $(m,t,0, \mathsf{d})$-share, and source 2 shares $\mathbf{B}$ in the form of $(m,t,0,\mathsf{r})$-share, with the workers. In other words,  worker $n$ has access to $[\Ab]_n \defeq \Ab(\alpha_n)$ and $[\Bb]_n \defeq \Bb(\alpha_n)$, for some distinct $\alpha_n \in \mathbb{F}$, where     $\Ab(x) \defeq \mathbf{F}^{(\mathsf{d})}_{\mathbf{A},m,t,0}(x)$ and $\Bb (x) \defeq \mathbf{F}^{(\mathsf{r})}_{\mathbf{B},m,t,0}(x)$, for all $n \in [N]$. 
  
 \textbf{2- Subsharing phase:}
 In this phase worker $n$ subshares its shares as follows:  
 \begin{itemize} 
 \item Worker $n$ shares $[\Ab]_n$, using a $(1, t, m-1,   \mathsf{d})$-sharing, with other workers. In other words, it forms a polynomial as
 \begin{align}
     \mathbf{A}_n(x) & \defeq  \mathbf{F}^{(\mathsf{d})}_{[\Ab]_n,1,t, m-1}(x)=   [\Ab]_n+\displaystyle\sum_{k=1}^{t}\mathbf{R}_k^{(n)}x^{m+k-1}, 
\end{align}
where  $\mathbf{R}_k^{(n)}$, $n \in [N]$, $k \in [t]$, are chosen independently and uniformly at random from $\mathbb{F}^{z \times \frac{z}{m}}$. Then, it  
 sends $ \mathbf{A}_n(\alpha_{n'})$ to worker ${n'}$, for all $n' \in [N]$.

\item To subshare $[\Bb]_n$, worker $n$ follows a different scheme. Worker $n$ first partitions $[\Bb]_n$ into $m$ equal-size matrices as 

 \begin{align}
  [\Bb]_n=
 \begin{bmatrix}
    &[\Bb]_{n,0}\\
    &[\Bb]_{n,1}\\
    &\vdots\\
    &[\Bb]_{n,m-1}
 \end{bmatrix}.
 \end{align}

Worker $n$ shares $ [\Bb]_{n,j}$ in the form of $(1, t, m-j-1,   \mathsf{d})$-share, with other workers. In other words, it forms
 \begin{align}
     \mathbf{B}_n^{(j)}(x) 
      \defeq 
     \mathbf{F}^{(\mathsf{d})}_{
     [\Bb]_{n,j} ,1,t, m-j-1 } (x)=  [\Bb]_{n,j} +\displaystyle\sum_{k=1}^{t}\mathbf{R}_k^{(n,j)}x^{m-j+k-1},
\end{align}
 where $\mathbf{R}_k^{(n,j)}$, $n \in [N]$, $j \in \{0,1,\dots,m-1\}$, $k \in [t]$,  are chosen independently and uniformly at random from $\mathbb{F}^{\frac{z}{m} \times \frac{z}{m}}$ , and sends   $\mathbf{B}_n^{(j)}(\alpha_{n'})$ to worker ${n'}$, for all $n' \in [N]$.  

 \end{itemize}

 Let us define $\mathbf{B}_n(x)$ as
\begin{align}
\mathbf{B}_n(x) \defeq\displaystyle\sum_{j=0}^{m-1}\mathbf{B}_n^{(j)}(x)x^j =& \sum_{j=0}^{m-1} x^j \left(   [\Bb]_{n,j} +\displaystyle\sum_{k=1}^{t}\mathbf{R}_k^{(n,j)}x^{m-j+k-1} \right)
 \\=&
 \sum_{j=0}^{m-1}   [\Bb]_{n,j} x^j +\displaystyle\sum_{k=1}^{t} x^{m+k-1}  \sum_{j=0}^{m-1} \mathbf{R}_k^{(n,j)}
 \\=& \sum_{j=0}^{m-1}   [\Bb]_{n,j} x^j +\displaystyle\sum_{k=1}^{t} x^{m+k-1}  \tilde{\mathbf{R}}_k^{(n)},
    \end{align}
where   $\tilde{\mathbf{R}}_k^{(n)} \defeq  \displaystyle\sum_{j=0}^{m-1} \mathbf{R}_k^{(n,j)}$. Thus, $\mathbf{B}^T_n(x)$ is indeed forms an $(m,t,0,\mathsf{d})$-share of $[\Bb]_n^T$. Finally, each worker ${n'}$ computes $\mathbf{B}_n(\alpha_{n'})= \displaystyle\sum_{j=0}^{m-1}\alpha_{n'}^j\mathbf{B}_n^{(j)}(\alpha_{n'})$.
   
\begin{Remark}
Note that worker $n$ could have directly send $\mathbf{B}_n(\alpha_{n'})$ to worker $n'$, rather than sending  $\mathbf{B}_n^{(j)}(\alpha_{n'})$, $j =0,\ldots, m-1$. This two-step approach will be exploited later to deal with adversaria
\end{Remark}

\textbf{3- Constructing the product of subshares:} Up to here, each worker $n$ has formed two polynomials $\Ab_n(x)$ and $\Bb_n(x)$, and sends $\Ab_n(\alpha_{n'})$ and $\Bb_n(\alpha_{n'})$ to worker ${n'}$, for all ${n'} \in [N]$. In this phase, worker $n$ follows a certain procedure to form 
$\mathbf{F}^{(\mathsf{d})}_{[\Ab]_n [\Bb]_n^T,m,t,0}(x)$, and sends $\mathbf{F}^{(\mathsf{d})}_{[\Ab]_n [\Bb]_n^T,m,t,0}(\alpha_{n'})$ to worker $n'$, $n' \in [N]$. We note that since $[\Ab]_n$ and $[\Bb]_n^T$ are available at worker $n$ (from Phase 1), this worker can directly form a polynomial in the form of $\mathbf{F}^{(\mathsf{d})}_{[\Ab]_n [\Bb]_n^T,m,t,0}(x)$. However, for the reasons that will be explained later, and to fight with adversarial nodes, 
worker $n$ uses  $\Ab_n(x)$ and $\Bb_n(x)$ to develop $\mathbf{F}^{(\mathsf{d})}_{[\Ab]_n [\Bb]_n^T,m,t,0}(x)$, by following an indirect approach. 
For this purpose, worker $n, n \in [N]$, can find some polynomials $\mathbf{O}_0^{(n)}(x),\mathbf{O}_1^{(n)}(x),\dots,\mathbf{O}_{m+t-2}^{(n)}(x)$ of degree $t$ such that 
\begin{align}
\label{eq4}
 \Cb_n(x)=\mathbf{A}_n(x)\mathbf{B}^T_n(x)-\displaystyle\sum_{\ell=0}^{m+t-2}x^{m+\ell}\mathbf{O}^{(n)}_\ell(x), 
\end{align}
forms a polynomial of degree $m+t-1$. One can see that $\Cb_n(x)$ is in the form of $\mathbf{F}^{(\mathsf{d})}_{\mathbf{A}(\alpha_n)\mathbf{B}^T(\alpha_n),m,t,0}(x)$. In what follows, we first show that it is possible to construct polynomials $\mathbf{O}_0^{(n)}(x),\mathbf{O}_1^{(n)}(x),\dots,\mathbf{O}_{m+t-2}^{(n)}(x)$ of degree $t$ such that deg($\Cb_n(x)$)=$m+t-1$.

Let us partition $[\Ab]_n[\Bb]_n^T$ into $m$ equal size matrices as 
\begin{align}
[\Ab]_n[\Bb]_n^T=[ \{[\Ab]_n[\Bb]_n^T\}_1, \{[\Ab]_n[\Bb]_n^T\}_2, \ldots, \{[\Ab]_n[\Bb]_n^T\}_{m-1}].
\end{align}
First, one can see  that the coefficient of $x^j$  in \eqref{eq4} is $\{[\Ab]_n[\Bb]_n^T\}_j$ for all $j$, $0 \leq j \leq m-1$. This  can be verified by calculating $\mathbf{A}_n(x)\mathbf{B}_n(x)$ directly and in each polynomial $x^{m+\ell} \mathbf{O}^{(n)}_\ell(x)$,  $0 \leq \ell \leq m+t-2$, the degree of each term is at least $m$.

To construct $\mathbf{O}_\ell^{(n)}(x)$, we use the following scheme. Let us show $\mathbf{A}_n(x)\mathbf{B}_n(x)$ as follows.
\begin{align}
\mathbf{A}_n(x)\mathbf{B}^T_n(x)= \mathbf{D}_0^{(n)}+\mathbf{D}_1^{(n)}x+\mathbf{D}_2^{(n)}x^2+\dots+\mathbf{D}_{2t+2m-2}^{(n)}x^{2t+2m-2}.
   \label{basteAB}
\end{align}
Then, worker $n$ chooses $\mathbf{O}_{m+t-2}^{(n)}(x)$ as
\begin{align*}
    \mathbf{O}_{m+t-2}^{(n)}(x)= \mathbf{R}_{m+t-2,0}&+\mathbf{R}_{m+t-2,1}x+ \mathbf{R}_{m+t-2,t-1}x^{t-1}+\mathbf{D}^{(n)}_{2t+2m-2}x^t,
\end{align*}
where $\mathbf{R}_{m+t-2,0}, \ldots, \mathbf{R}_{m+t-2,t-1}$ are chosen independently and uniformly at random from $\mathbb{F}^{z \times \frac{z}{m}}$. This makes us sure that in $\Cb_n(x)$ (see \eqref{eq4}), the coefficient of $x^{2m+2t-2}$ is zero. 

Then worker $n$ chooses $\mathbf{O}_{m+t-3}^{(n)}(x)$, as 
\begin{align*}
\mathbf{O}_{m+t-3}^{(n)}(x)= \mathbf{R}_{m+t-3,0}+\mathbf{R}_{m+t-3,1}x+\dots
+\mathbf{R}_{m+t-3,t-1}x^{t-1}+(\mathbf{D}^{(n)}_{2t+2m-3}-\mathbf{R}_{m+t-2,t-1})x^t,
\end{align*}
where $\mathbf{R}_{m+t-3,0}, \ldots, \mathbf{R}_{m+t-3,t-1}x^{t-1}$ are chosen are independently and uniformly at random from $\mathbb{F}^{z \times \frac{z}{m}}$. This makes us sure that in $\Cb_n(x)$ (see \eqref{eq4}), the coefficient of $x^{2m+2t-3}$ is zero. Using the same procedure recursively, we can find other polynomials as follows.

\begin{align*}
    \mathbf{O}_{m+t-4}^{(n)}(x)= \mathbf{R}_{m+t-4,0}+\mathbf{R}_{m+t-4,1}x+\dots &+ \mathbf{R}_{m+t-4,t-1}x^{t-1}\\&+(\mathbf{D}^{(n)}_{2t+2m-4}-\mathbf{R}_{m+t-3,t-1}-\mathbf{R}_{m+t-2,t-2})x^t,\\
    &\vdots\\
    \mathbf{O}_{0}^{(n)}(x)= \mathbf{R}_{0,0}+\mathbf{R}_{0,1}x+\dots+\mathbf{R}_{0,t-1}x^{t-1}&+ (\mathbf{D}^{(n)}_{t+1}-\mathbf{R}_{1,t-1}-\dots-\mathbf{R}_{t,0})x^t,
\end{align*}
where $\mathbf{R}_{i,j}$ are chosen independently and uniformly at random  from $\mathbb{F}^{z \times \frac{z}{m}}$ and $\mathbf{D}^{(n)}_{\ell}(x)$ are the coefficients of the expansion of  $\mathbf{A}_n(x)\mathbf{B}^T_n(x)$ shown in \eqref{basteAB}.

One can easily verify that with the above choices of $\mathbf{O}_0^{(n)}(x),\mathbf{O}_1^{(n)}(x),\dots,\mathbf{O}_{m+t-2}^{(n)}(x)$, then $\Cb_n(x)$ is of degree $m+t-1$, and it can be written  as
\begin{align}
\label{Cnx}
   & \mathbf{C}_n(x)= \displaystyle\sum_{\ell=0}^{m-1} \{[\Ab]_n[\Bb]_n^T \}_\ell x^\ell+\displaystyle\sum_{k=1}^{t}\mathbf{Z}_k^{(n)}x^{m+k-1}, 
\end{align} 
where $\mathbf{Z}_k^{(n)}$ has uniform distribution in $\mathbb{F}^{z \times \frac{z}{m}}$, independent from $[\mathbf{A}]_n[\mathbf{B}^{T}]_n$. Thus  $\mathbf{C}_n(x)$
is in the form of $\mathbf{F}^{(\mathsf{d})}_{[\mathbf{A}]_n[\mathbf{B}]_n^T,m,t,0}(x)$.

Then worker $n$ sends $\mathbf{O}_0^{(n)}(\alpha_{n'}),\mathbf{O}_1^{(n)}(\alpha_{n'}),\dots,\mathbf{O}_{m+t-2}^{(n)}(\alpha_{n'})$ to worker ${n'}$, ${n'} \in N$. 
Thus each worker $n'$ can form  $\mathbf{C}_n(\alpha_n')$ using \eqref{eq4}, and using the fact that it has access to $\Ab_n(\alpha_{n'})$ and $\Bb_n(\alpha_{n'})$ from Phase 2 and $\mathbf{O}_0^{(n)}(\alpha_{n'}),\mathbf{O}_1^{(n)}(\alpha_{n'}),\dots,$ $\mathbf{O}_{m+t-2}^{(n)}(\alpha_{n'})$ from this phase.

\textbf{4- Constructing shares of product:} 
First we need to express an important theorem from \cite{nodehi2018entangled}.
\begin{theorem}
\label{combination lagrange}
Consider an arbitrary polynomial $\mathbf{F}(x)= \displaystyle\sum^{k}_{i=0}\mathbf{C}_i x^i$ of degree $k$. Assume that we have the value of $\mathbf{F}(x)$ at some points $a_1,a_2,\dots,a_M \in \mathbb{F}$. We can write any coefficient of polynomial $\mathbf{F}(x)$ as a linear combination of $\mathbf{F}(a_1),\mathbf{F}(a_2),\dots,\mathbf{F}(a_n)$ if and only if $M \geq k+1$.
\end{theorem}

As it was shown in Theorem \ref{combination lagrange}, if $N \geq 2m+2t-1$ there exist $\lambda_1,\lambda_2,...,\lambda_N \in \mathbb{F}$ such that
\begin{align}
\label{eqC}
    \displaystyle\sum_{n=1}^{N}\lambda_n [\mathbf{A}]_n [\mathbf{B}]^T_n=\mathbf{C}.
\end{align}

In this phase, each worker $n'$, $n'\in [N]$,  calculates  $\sum_{n=1}^N \lambda_n  \mathbf{C}_n(\alpha_n')$. 

We note that 
\begin{align*}
 \sum_{n=1}^N  \lambda_n \mathbf{C}_n(x)
 \overset{(a)}{=}&
\sum_{n=1}^N \lambda_n  \sum_{\ell=0}^{m-1} \{[\Ab]_n[\Bb]_n^T \}_\ell  x^\ell+   \sum_{n=1}^N   \lambda_n \sum_{k=1}^{t}\mathbf{Z}_k^{(n)}x^{m+k-1}\\
  = &   \sum_{\ell=0}^{m-1}  x^\ell  \sum_{n=1}^N \lambda_n \{[\Ab]_n[\Bb]_n^T \}_\ell  +  \sum_{k=1}^{t}  x^{m+k-1} \sum_{n=1}^N   \lambda_n  \mathbf{Z}_k^{(n)}\\
    \overset{(b)}{=}  &  \sum_{\ell=0}^{m-1}  x^\ell \Cb_\ell  +  \sum_{k=1}^{t}  x^{m+k-1}   \tilde{\mathbf{Z}}_k^{(n)}, 
\end{align*}
where (a) follows from \eqref{Cnx} and (b) follows from \eqref{eqC} and also $\Cb= \Ab \Bb^T=[\Cb_0, \ldots, \Cb_{m-1}]$. In the above equation  $\tilde{\mathbf{Z}}_k^{(n)} \defeq \sum_{n=1}^N   \lambda_n  \mathbf{Z}_k^{(n)}$. Thus, $\sum_{n=1}^N  \lambda_n \mathbf{C}_n(x)$ forms  $\mathbf{F}^{(\mathsf{d})}_{\mathbf{C},m,t,0}(x)$, and each worker $n$ has $\mathbf{F}^{(\mathsf{d})}_{\mathbf{C},m,t,0}(\alpha_n)$. This is what we were looking for. 

In the above algorithm, if we have adversarial nodes, then in each phase, an adversarial node $n$ can send wrong information to the other nodes and make the entire algorithm to fail.  To take into account the challenges that we face, when we have adversarial behavior and provide the solution which is resistance to this behavior, we need to introduce four subroutines which will be developed in the next section. Then in Section \ref{with}, we will propose modified version of the algorithm presented in this section, relying on subroutines that we introduce in Section \ref{preliminaries}.

\section{preliminaries}
\label{preliminaries}
In this section we extend some of the existing subroutines used in multi-party computation to be able to apply to the proposed scheme for computing an arbitrary polynomial of input matrices.

\subsection{Extended Verifiable Secret Sharing}
\label{EVSS}
A procedure to share a secret $s$ by a dealer, among some nodes or shareholders in such a way that any fewer than or equal to $t$ of colluding nodes cannot gain any information about secret $s$, while any set of honest nodes more than some threshold can recover the secret is called secret sharing. Secret sharing, first, was introduced by Shamir \cite{shamir1979share1} and Blakley \cite{blakley1979safeguarding}, independently, in 1979.

Secret sharing is a basic tool in cryptography and has been used in many applications such as e-voting schemes, crypto-currencies and access control systems. In Shamir secret sharing, the dealer uses the secret as the constant term of a polynomial of degree $t$, where the other coefficients are chosen randomly and uniformly from the field. Then the dealer sends the value of this polynomial at different points to different nodes, as their shares. It can be shown that, this scheme is information theoretically secure. In this scheme, we assume that the dealer is trusted and always sends consistent shares to the other nodes, i.e., it chooses points on a polynomial of degree $t$. In some applications, the dealer may be corrupted. In this case, we need a mechanism to be able to verify the consistency of the shares. Chor et al. \cite{chor1985verifiable} introduce verifiable secret sharing (VSS), which enables nodes to confirm whether their shares are consistent or not. The work of \cite{chor1985verifiable} has been followed by many other results, which can be categorized into two major approaches.

\begin{enumerate}
	\item Computational VSS scheme:
	In this case, we assume that adversaries have bounded computing power that limits their ability to solve some mathematical problems with extensive complexity, such as finding prime divisors of a large composite number. Some examples of computational VSS can be found in \cite{feldman1987practical,pedersen1991non}.
	\item Information theoretically secure VSS scheme:
	In this case, we do not limit the adversaries in term of computational power or storage size. Those kinds of systems are information-theoretically secure, i.e., the security holds, even if the adversary has unbounded computing power, such as \cite{benaloh1986secret,stinson1999unconditionally,patra2009efficient}.
\end{enumerate}

Here, we propose an \emph{extended verifiable secret sharing}, as an extended version of $\mathsf{VSS}$~\cite{chor1985verifiable}, that has the following properties:
\begin{enumerate}
	\item If the dealer is malicious, and the shares that it sends to the other nodes are not consistent, i.e., are not some points on a polynomial of specified degree, the honest nodes in collaboration with each other will realize that and reject the shares.
	\item If the dealer is honest, then the malicious workers cannot deceive the honest nodes and convince them that the dealer is malicious, and thus each honest node accepts its share.
\end{enumerate}
In its original form \cite{ben1988completeness}, to share a secret $s$ from a field $\mathbb{F}$, the dealer chooses a bivariate polynomial $S(x,y)$, uniformly at random from the set of all bivariate polynomials of degree $t$, with respect to each of the variables $x$ and $y$, with coefficients from $\mathbb{F}$, subject to $S(0,0)=s$. Then the dealer sends $S(x,\alpha_n)$ and $S(\alpha_n,y)$ to the worker $n$, for all $n \in [N]$ and some distinct $\alpha_n \in \mathbb{F}$. The redundancy in this scheme, allows honest workers to verify the consistency of shares through communication with other workers. 

We do not explain the algorithm of $\mathsf{VSS}$. The reader can read it in details in \cite{fullproof}. Here we explain the proposed scheme where in some special cases it reduces to the verifiable secret sharing. Briefly, the proposed scheme works as follows. The full protocol of $\mathsf{EVSS}$ is explained in details in Algorithm \ref{ProtocolEVSS}.

Let us assume that the dealer has a matrix $\mathbf{W}$, partitioned equally in $m$ sub-matrices,  as $\mathbf{W}= [\mathbf{W}_0,\mathbf{W}_1, \dots,\mathbf{W}_{m-1}]$. We also assume that the dealer forms a polynomial $\mathbf{Q}(z)\defeq\displaystyle\sum_{j=0}^{m-1} \mathbf{W}_{j}x^{j} +\sum_{k=1}^{t}\mathbf{R}_k x^{m+\delta+k-1}$ where $\mathbf{R}_k$, $k=0, \ldots t-1$,   are chosen independently and uniformly at random. It plans (or simply pretends) to send $\mathbf{Q}(\alpha_n)$ to node $n$. In a $({\mathbf{W}}, m,t, \delta)-{\mathsf{EVSS}}$ algorithm, the dealer embeds  the polynomial $\mathbf{Q}(z)$ into a bivariate polynomial $\mathbf{S}(x,y)$. This means that it forms  $\mathbf{S}(x,y)$ of degree $m+t+\delta-1$ with respect to each variable, such that $\mathbf{S}(0,z)=\mathbf{Q}(z)$. Other coefficients of $\mathbf{S}(x,y)$ are chosen independently and  uniformly at random. Then, the dealer sends worker $n$ two univariate polynomials $f_n(x)=\mathbf{S}(x,\alpha_n)$ and $g_n(y)=\mathbf{S}(\alpha_n,y)$ as intermediate shares, for all $n \in [N]$. This approach introduces some redundancy and allows the nodes for verification. One can see that $\forall n,n' \in [N], n \neq n'$, we have $ f_n(\alpha_{n'})=g_{n'}(\alpha_n)$. This property is used to verify the consistency of share as follows. Each pair of workers $(i,j)$, send the values $f_i(\alpha_j)$, $g_i(\alpha_j)$, $f_j(\alpha_i)$ and $g_j(\alpha_i)$ to each other. Then each of them can verify that either their univariate polynomials (intermediate shares) are \emph{pairwise consistent}, i.e. $f_i(\alpha_j)=g_j(\alpha_i)$ and $f_j(\alpha_i)=g_i(\alpha_j)$ or not. If it is not, e.g., worker $n$ receives values $f_{n'}(\alpha_{n})$ and $g_{n'}(\alpha_{n})$ from worker $n'$ and $f_n(\alpha_{n'}) \neq g_{n'}(\alpha_n)$ or $f_{n'}(\alpha_n) \neq g_n(\alpha_{n'})$, worker $n$ will broadcast a \texttt{complaint} massage including $(n,n',f_n(\alpha_{n'}),g_n(\alpha_{n'}))$. In this case, the dealer examines that if these values are correct, then reveals nothing, else it broadcasts the entire polynomials $f_n(x)$ and $g_n(y)$. Now all the other parties can verify that these new polynomials are consistent with their shares or not. If it is, they broadcast \texttt{consistent}. There must be at least $N-t$ workers which broadcast \texttt{consistent} to accept that the dealer is not malicious and the polynomials are correct.

As we shall see in Algorithm \ref{ProtocolEVSS}, this method prevents the dealer to distribute non-consistent shares. Finally each node $n$ derives $f_n(0)=\mathbf{S}(0,\alpha_n)=\mathbf{Q}(\alpha_n)$ as its polynomial share.

\begin{Remark}
Whenever the dealer constructs $\mathbf{S}(x,y)$ such that, $\mathbf{S}(0,z)=\mathbf{Q}(z)$ and shares $\mathbf{S}(x,y)$ using the following $\mathsf{EVSS}$ scheme, we say that the dealer shares $\mathbf{Q}(z)$.
\end{Remark}

\begin{Remark}
	For some special cases, $\mathsf{EVSS}$ reduces to the conventional $\mathsf{VSS}$. In particular $(s,1,t,0)-\mathsf{EVSS}$ is the same as $\mathsf{VSS}$ to share $s$.
	
\end{Remark}

\begin{theorem}
	\label{consistency}
	Let $\mathcal{K} \subset [N]$ be a set of integers and $\{f_k(x),g_k(y)\}_{k \in \mathcal{K}}$ be a set of pairs of polynomials of degree $m'$ where $|\mathcal{K}| \geq m'+1$. And let $\{\alpha_k\}_{k \in \mathcal{K}}$ be some distinct non-zero elements in $\mathbb{F}$. If for every $i,j \in \mathcal{K}$, $f_i(\alpha_j)=g_j(\alpha_i)$, then there exists a unique bivariate polynomial $S(x,y)$ of degree $m'$ with respect to each of the variables such that $f_k(x)=S(x,\alpha_k)$ and $g_k(y)=S(\alpha_k,y)$ for every $k \in \mathcal{K}$.
\end{theorem}
Proof: See \cite{fullproof}.

\begin{Remark}
The above theorem implies that, if in an $\mathsf{EVSS}$ invocation there exists some honest workers $i_1,i_2,\dots,i_{m+\delta+t}$ such that for each pair of them like $i_u$ and $i_v$, their shares are consistent, i.e., $f_{i_u}(\alpha_{i_v})=g_{i_v}(\alpha_{i_u})$ and $f_{i_v}(\alpha_{i_u})=g_{i_u}(\alpha_{i_v})$, then there exists exactly one bivariate polynomial $S(x,y)$ such that $S(x,\alpha_{i_l})=f_{i_l}(\alpha_{i_l})$ and $S(\alpha_{i_l},x)=g_{i_l}(\alpha_{i_l})$, for all $l \in [m+\delta+t]$.
\end{Remark}

\begin{theorem}
	\begin{enumerate} Assume that $N \geq 3t +m$.
		\item If the dealer is honest and follows Algorithm \ref{ProtocolEVSS} to invoke $(\mathbf{W},p,m,t)-\mathsf{EVSS}$ using $S(x,y)$ such that $S(0,z)=Q(z)$, then each honest worker $i$, who follows Algorithm \ref{ProtocolEVSS}, will accept $f_i(0)=Q(\alpha_i)$, as its share.
		\item If the dealer is malicious, either it follows Algorithm \ref{ProtocolEVSS} and each honest node $i$, correctly admits $f_i(0)$ as its verified share of the secret, or all of them reject it, and output $\perp$.
		
	\end{enumerate}
	
\end{theorem}

\textbf{Proof:}
\begin{enumerate}
	\item First, consider the case where the dealer is honest, i.e., it follows Algorithm \ref{ProtocolEVSS}. Thus for any $i$,$j$ in the set of all honest workers, $f_i(\alpha_j)=g_j(\alpha_i) \text{ and }g_i(\alpha_j)=f_j(\alpha_i)$.
	
	Assume that node $i$ is honest and node $j$ is corrupted. If the corrupted worker $j$, in Step 2 of Algorithm \ref{ProtocolEVSS}, sends incorrect values instead of $(f_j(\alpha_i),g_j(\alpha_i))$ to the honest worker $i$, then the honest node $i$, broadcasts a \texttt{complaint} message as explained in Step 3 of Algorithm \ref{ProtocolEVSS}. Recall that the \texttt{complaint} message includes $(i,j,f_i(\alpha_j),g_i(\alpha_j))$, where $f_i(\alpha_j)$ and $g_i(\alpha_j)$  are the correct values of $g_j(\alpha_i)$ and $f_j(\alpha_i)$. However node $j$, which is corrupted, already has access to these values, therefore broadcasting this information by node $i$ does not reveal anything new.
	On top of that, since these values are correct, the dealer will not reveal the polynomials of the honest worker, i.e., $f_i(x)$ and $g_i(x)$.
	
	On the other hand, if a malicious worker sends a \texttt{complaint} message, including incorrect values, then the honest dealer will reveal its polynomials. Note that these polynomials have been revealed to an adversary before, and broadcasting it by the dealer does not reveal anything new. When this case occurs, all honest parties verify consistency of revealed polynomials and broadcast \texttt{consistent}. Thus at least $N-t$ workers broadcast \texttt{consistent} message in Step 5. Finally, each honest node $i$ accepts $f_i(0)$ as its share, according to Algorithm \ref{ProtocolEVSS}.
	
	\item Next, assume that the dealer is malicious. In this case, two honest workers $i \text{ and } j$ may receive polynomials that are not consistent with each other, i.e., $f_i(\alpha_j) \neq g_j(\alpha_i) \text{ or }g_i(\alpha_j) \neq f_j(\alpha_i)$. In this case, since both of the honest workers broadcast conflicting \texttt{complaint} messages, then the dealer is forced to response by at least a valid  \texttt{reveal} message containing of shares of worker $i$ or $j$ in Step 4 of Algorithm \ref{ProtocolEVSS}. If all the revealed messages are consistent with intermediate shares of other honest workers, then they broadcast \texttt{consistent} messages in Step 5. Else, they will not broadcast \texttt{consistent}. There must be at least $N-t$ workers which broadcast \texttt{consistent} to accept that the shares are consistent, else, all honest nodes output $\perp$. If there are at least $N-t$ \texttt{consistence} messages, then there are at least $(N-t)-t$ honest workers among who broadcast \texttt{consistent} in Step 5 (Among at least $N-t$ workers that broadcast \texttt{consistent}, there are at most $t$ malicious workers, thus there are  at least $N-t-t$ honest nodes among them). If $N \geq 3t+m$, then it implies that there exist at least $m+t$ honest workers, such that their shares are \emph{pairwise consistent} with each other. Therefore as a result of Theorem \ref{consistency}, their polynomials are received from a unique bivariate polynomial $S(x,y)$ of degree $m+t-1$. Assume that there exists an honest worker $k$, such that its polynomial, say $f_k(x)$ is not equal with $S(x,\alpha_k)$. Since at least $m+t$ honest workers verify that $f_k(\alpha_j)= S(\alpha_j,\alpha_k)$ and deg($f_k(x)$)$\leq m+t-1$, then we can conclude that the polynomial $f_k(x)$ is unique and obtained from $S(x,y)$. Thus $f_k(x)$ is consistent with other shares, and each honest node $i$ admits $f_i(0)$ as a verified share of the secret.
\end{enumerate}

The procedure in Algorithm \ref{ProtocolEVSS} is similar to the verifiable secret sharing algorithm in \cite{asharov2017full} except that in first step we may put data in more than one coefficient. To capture that coding more than one matrices simultaneously in just one polynomial. As a result of that, the threshold here is different from  the threshold of the conventional $\mathsf{VSS}$.

\newpage

\begin{Algorithm}{Extended Verifiable Secret Sharing}
\label{ProtocolEVSS}

\textit{Initializing:} The dealer $D=P_0$ holds a secret input $\mathbf{W}=[\mathbf{W}_0,\mathbf{W}_1,\dots,\mathbf{W}_{m-1}]$, where $\mathbf{W}_j \in \mathbb{F}^{e \times r}$, for all $j \in \{0,1,\dots,m-1\}$. The field $\mathbb{F}$ and $N$ non-zero elements $\alpha_1,\alpha_2,\dots,\alpha_{N} \in \mathbb{F}$ are known to all nodes.

\textit{The protocol:}
\begin{enumerate}
	\item \textbf{Sending shares by the dealer:}
	\begin{enumerate}
		\item
		The dealer forms a bivariate polynomial $\mathbf{S}(x,y)=\displaystyle\sum_{j=0}^{m+t-1}\displaystyle\sum_{i=0}^{m+t-1}\mathbf{S}_{i,j}x^iy^j$ such that $\mathbf{S}_{0,j}=\mathbf{W}_j$ for all $0 \leq j \leq m-1$ and $\mathbf{S}_{i,j},$ are selected independently and uniformly at random from $\mathbb{F}^{e \times r}$, for all $i \neq 0 \text{ or } m \leq j$ .
		
		\item
		For every $i \in [N]$, the dealer sends $f_i(x) \defeq \boldsymbol{S}(x,\alpha_i)$ and $g_i(y) \defeq \boldsymbol{S}(\alpha_i,y)$ to each party $P_i$.
		
	\end{enumerate}
	
	\item \textbf{storing and exchanging subshares by each party $P_i$:}
	\begin{enumerate}
		\item
		Party $P_i$ keeps the polynomials $f_i(x)$ and $g_i(y)$, received from the dealer.
		
		\item
		Party $P_i$ sends $f_i(\alpha_j)$ and $g_i(\alpha_j)$ to party $P_j$, $j \in [N]$.
		
	\end{enumerate}
	
	\item \textbf{broadcasting complaints by each party $P_i$:}
	\begin{enumerate}
		\item
		Party $P_i$ stores $u_j \defeq f_j(\alpha_i)$ and $v_j \defeq g_j(\alpha_i)$, where $f_j(\alpha_i)$ and $g_j(\alpha_i)$ are received from party $P_j$. If $u_j\neq g_i(\alpha_j)$ or $v_j \neq f_i(\alpha_j)$, party $P_i$ broadcasts $\texttt{complaint}(i,j,f_i(\alpha_j),g_i(\alpha_j))$.
		
		\item
		If no \texttt{complaint} is broadcasted by any of the parties, then party $P_i$ stores $f_i(0)$ as the verified share and halts.
		
	\end{enumerate}
	\item \textbf{resolving complaints by the dealer:} For every \texttt{complaint} message received, the dealer does the following:
	\begin{enumerate}
		\item
		When the dealer receives a complaint message $\texttt{complaint}(i,j,u,v)$ which is broadcasted by $P_i$, it verifies that $u=\boldsymbol{S}(\alpha_j,\alpha_i)$ and $v=\boldsymbol{S}(\alpha_i,\alpha_j)$. If at least one of these equations is incorrect, then it broadcasts $\texttt{reveal}(i,f_i(x),g_i(y))$, otherwise, it does nothing.
	\end{enumerate}
	\item \textbf{evaluating complaint resolutions by each party $P_i$:}
	\begin{enumerate}
		\item
		$(j,k)$ is known as \texttt{joint complaint} if two complaint messages, $\texttt{complaint}(k,j,u_1,v_1)$ and $\texttt{complaint}(j,k,u_2,v_2)$ broadcasted simultaneously  by $P_k$ and $P_j$, respectively. If in the response of \texttt{joint complaint}$(j,k)$ the dealer broadcasts at least one of $\texttt{reveal}(k,f_k(x)$ $,g_k(y))$ or $\texttt{reveal}(j,f_j(x),g_j(y))$, then party $P_i$ goes to the next step otherwise, it goes to Step 6.
		\item
		First party $P_i$ considers all \texttt{reveal} messages sent by the dealer:
		\begin{enumerate}
			\item 
			If there exist a \texttt{reveal}$(j,f_j(x),g_j(y))$ message such that $j=i$, then party $P_i$ replaces the new polynomials $f_j(x) \text{ and } g_j(y)$ with the stored ones. Then it goes to Step 6.
			\item
			If there exists a \texttt{reveal} message in the set which was not consistent with $P_i$ shares, i.e., there exist \texttt{reveal}\\$(j,f_j(x),g_j(y))$ with $j \neq i$ such that $f_i(\alpha_j) \neq g_j(\alpha_i)$ or $g_i(\alpha_j) \neq f_j(\alpha_i)$, then it goes to Step 6.
		\end{enumerate}
		For each party $P_i$, if there is not a message in the set that satisfies at least one of the above conditions, then it broadcasts \texttt{consistent}. 
		\item
		Party $P_i$ broadcasts the message \texttt{consistent}.
	\end{enumerate}
	\item \textbf{if there is complaints - each party $P_i$:} If at least $N-t$ parties broadcast \texttt{consistent}, each party $P_i$ outputs $f_i(0)$. Otherwise, it outputs $\perp$.
\end{enumerate}
\end{Algorithm}

\newpage

\subsection{Multiplication of Shares By a Matrix}
\label{sharesmultmatrix}
Assume that each worker $n$ has a private value $x_n$ and all parties need to know the product of $\mathbf{x}=[x_1,x_2,\dots,x_N]$ and an arbitrary matrix $\mathbf{H} \in \mathbb{F}^{N\times m}$ where the privacy constraints is preserved. It means that each worker $n$ learns nothing about $x_{n'}, n' \neq n, n' \in [N]$ beyond $\mathbf{x}\mathbf{H}$. The privacy constraints can be written formally similar to what stated in the problem formulation in Section \ref{sec:Problem Setting}. This scheme is called \emph{multiplication of shares by a matrix} which is explained in \cite{fullproof}. To be self-contained, here we review the scheme.

To do this, first node $n$ uses $(x_n,1,t,\delta)-\mathsf{EVSS}$ where $\delta \in \mathbb{N}$, to share $x_n$, using a bivariate polynomial denoted by $\mathbf{S}_n(x,y)$. We define $\mathbf{Q}_n(z)=\mathbf{S}_n(0,z)$. According to EVSS, $\mathbf{Q}_n(0)=x_n$. For now, assume that in this subsection, each node $n$ shares the correct value of $x_n$. Later, in Subsection \ref{subshare}, we will discuss the case where some of the nodes do not share the correct values.
Then each worker $n'$ has access to $Q_1(\alpha_{n'}),Q_2(\alpha_{n'}),\dots,Q_N(\alpha_{n'})$ for all $n' \in [N]$. Node $n'$ computes $[Q_1(\alpha_{n'}),Q_2(\alpha_{n'}),\dots$ $,Q_N(\alpha_{n'})]\mathbf{H} \defeq [y_{n',1},y_{n',2},\dots,y_{n',m}]$, and sends $[\tilde{y}_{n',1},\tilde{y}_{n',2},\dots,\tilde{y}_{n',m}]$ to node $n$ for all $n \in [N]$, where if Node $n'$ is honest  $[\tilde{y}_{n',1},\tilde{y}_{n',2},\dots,\tilde{y}_{n',m}]$ is equal to $[y_{n',1},y_{n',2},\dots,y_{n',m}]$, otherwise, it can be anything. Different nodes may receive different vector from node $n'$, if it is corrupted. Let us define matrix $\mathbf{Y}$ and $\mathbf{Q}$ as follows.

\begin{align}
\mathbf{Y}\defeq
\begin{bmatrix}
y_{1,1} &y_{1,2}  &\dots & y_{1,m}\\
y_{2,1} &y_{2,2}  &\dots & y_{2,m}\\
y_{3,1} &y_{3,2} &\dots& y_{3,m}\\
\vdots & \vdots & \ddots & \vdots \\
y_{N,1} &y_{N,2} &\dots & y_{N,m}\\
\end{bmatrix},    
\end{align}

\begin{align}
\mathbf{Q}\defeq
\begin{bmatrix}
\mathbf{Q}_1(\alpha_1)&\mathbf{Q}_2(\alpha_2)  &\dots & \mathbf{Q}_N(\alpha_1)\\
\mathbf{Q}_1(\alpha_2)&\mathbf{Q}_2(\alpha_2)  &\dots & \mathbf{Q}_N(\alpha_2)\\
\vdots & \vdots & \ddots & \vdots \\
\mathbf{Q}_1(\alpha_N)&\mathbf{Q}_2(\alpha_N)  &\dots & \mathbf{Q}_N(\alpha_N)\\
\end{bmatrix}  .
\end{align}
Thus
\begin{align}
\label{y=qh}
\mathbf{Y}=\mathbf{Q}\mathbf{H}.
\end{align}

Ideally, each node has matrix $\mathbf{Y}$, however, because of adversarial behaviour, in reality each node has $\mathbf{\Tilde{Y}}$ as follows.

\begin{align*}
\mathbf{\Tilde{Y}}\defeq
\begin{bmatrix}
\tilde{y}_{1,1} &\tilde{y}_{1,2}  &\dots & \tilde{y}_{1,m}\\
\tilde{y}_{2,1} &\tilde{y}_{2,2}  &\dots & \tilde{y}_{2,m}\\
\tilde{y}_{3,1} &\tilde{y}_{3,2} &\dots& \tilde{y}_{3,m}\\
\vdots & \vdots & \ddots & \vdots \\
\tilde{y}_{N,1} &\tilde{y}_{N,2} &\dots & \tilde{y}_{N,m}\\
\end{bmatrix}    
\end{align*}

For each honest node $i$, $i^{th}$ row of matrix $\mathbf{\Tilde{Y}}$ is equal to $i^{th}$ row of matrix $\mathbf{Y}$. Let us define $\mathbf{Y}(x)=[\mathbf{Y}_1(x),\mathbf{Y}_2(x),\dots,\mathbf{Y}_m(x)]\defeq[Q_1(x),Q_2(x),\dots,Q_N(x)]\mathbf{H}$. Obviously, $\mathbf{Y}_i(x)$ is of degree $k \defeq t+\delta$. Column $i$ of matrix $\mathbf{Y}$, consisting of $N$ points on a polynomial of degree $k$, is a Reed-Solomon Code \cite{wicker1999reed} of $\mathbf{Y}_i(x)$. Reed-Solomon Codes are a kind of \emph{Maximum Distance Separable} codes and meets Singleton bound \cite{cover2012elements}. Thus by using Reed-Solomon decoding procedure, workers can correct up to $\frac{N-k}{2}$ errors in each column. Since the number of malicious adversaries is at most $t$, so we need to have $\frac{N-k}{2} > t$. Note that $N \geq 2t+k=3t+\delta$ and in our procedure $\delta$ is less than $m$. Thus, it is sufficient that $N \geq 3t+m$. If $N \geq 3t+m$ each worker can recover $\mathbf{Y}(x)=[\mathbf{Y}_1(x),\mathbf{Y}_2(x),\dots,\mathbf{Y}_m(x)]=[Q_1(x),Q_2(x),\dots,Q_N(x)]\mathbf{H}$. So it can compute $[Q_1(0),Q_2(0),\dots,Q_N(0)]\mathbf{H}$ which is equal to $\mathbf{x}^T\mathbf{H}$. 
This procedure can be found in Algorithm \ref{protocolMat}, and is used in Subsection \ref{subshare} to verify that each node shares the correct private input. The proof of the privacy of the procedure is postponed to Section \ref{security}.

\newpage

\begin{Algorithm}{ Multiplication of Shares By a Matrix}
\label{protocolMat}
\textit{Initializing:} Each node $n$ holds a polynomial $Q_n(x)$, where $Q_n(0)=x_n$. A field $\mathbb{F}$, a matrix $\boldsymbol{H} \in \mathbb{F}^{N \times m}$, and $n$ non-zero elements $\alpha_1,\alpha_2,\dots,\alpha_{N} \in \mathbb{F}$ are known to all nodes.

\textit{The protocol:}
\begin{enumerate}
	\item Each node $n$ sends $Q_n(\alpha_{n'})$ to node $n'$ for all $n,n' \in [N]$, using $(x_n,1,t,0)-\mathsf{EVSS}$.
	
	\item Each node $n'$ stores $\mathbf{Q}(\alpha_{n'})\defeq [Q_1(\alpha_{n'}),Q_2(\alpha_{n'}),\dots,Q_N(\alpha_{n'})]$. If any of these entities is equal to $\perp$, it will be replaced with $0$.
	
	\item
	Each honest node $n$ computes $\mathbf{Y}(\alpha_n) \defeq \mathbf{Q}(\alpha_n)\mathbf{H}$ and then sends $\mathbf{Y}(\alpha_n)$ to node $n'$ for all $n,n' \in [N]$.
	
	\item
	Node $n'$ received $\tilde{\mathbf{Y}}(\alpha_n)$ from node $n$ (If node $n$ is honest, $\tilde{\mathbf{Y}}(\alpha_n)$ is equal to $\mathbf{Y}(\alpha_n)$, else, it is an arbitrary vector). 
	
	\item Each node $n'$ constructs matrix $\tilde{\mathbf{Y}}_{n'} \defeq [\tilde{\mathbf{Y}}(\alpha_1),\tilde{\mathbf{Y}}(\alpha_2),\dots,\tilde{\mathbf{Y}}(\alpha_N)]^T$. It uses Reed-Solomon decoding on each column of matrix $\tilde{\mathbf{Y}}_{n'}$ to recover vector $[\mathbf{Y}(\alpha_1),\mathbf{Y}(\alpha_2),\dots,\mathbf{Y}(\alpha_N)]$, and then computes
	$\mathbf{Y}(x)=[Q_1(x),Q_2(x),\dots,Q_N(x)]\mathbf{H}$, which enables it to compute $\mathbf{Y}(0) \defeq \mathbf{Q}(0)\mathbf{H}=[x_1,x_2,\dots,x_N]\mathbf{H}$.
	
\end{enumerate}

\end{Algorithm}

\newpage

\subsection{Subshare of shares in the form of $(\mathbf{F}(\alpha_n),1,t,\zeta-1)-\mathsf{EVSS}$}
\label{subshare}

Consider a polynomial $\mathbf{F}(x)$ of degree $p$ where $p\geq t$. Assume that each node $n$ has access to $\mathbf{F}(\alpha_n)$ as its private share. The goal is that each node $n$ is delivered a share of $\mathbf{F}(\alpha_n')$ from node $n'$, i.e, node $n$ has access to $\mathbf{Q}_1(\alpha_n),\mathbf{Q}_2(\alpha_n),\dots,\mathbf{Q}_N(\alpha_n)$, such that $\mathbf{Q}_{n'}(x) = \mathbf{F}(\alpha_{n'})+\mathbf{R}_{1,n'}x^{\zeta}+\mathbf{R}_{2,n'}x^{\zeta+1}+\dots+\mathbf{R}_{t,n'}x^{\zeta+t-1}$, where $\mathbf{R}_{1,n'},\mathbf{R}_{2,n'},\dots,\mathbf{R}_{t,n'}$ are chosen independently and uniformly at random, and $\zeta$ is some integer, for all $n,n' \in [N]$. In this system, at most $t$ of the nodes are malicious adversary. Our goal is to make sure that the sharing is done correctly, otherwise, the honest nodes identify the malicious nodes and omit them from the remaining part of computations. We will show that if the total number of nodes is grater than $2t+p$, we can reach to this goal.

 Assume that each node $n$ holds a value $\mathbf{F}(\alpha_n)$ as its private share, which is sampled from a polynomial $\mathbf{F}(x)=\displaystyle\sum_{i=0}^{p}\eta_i x^i$. In this subsection, the aim is that each worker re-shares its share with all other nodes, using $(\mathbf{F}(\alpha_n),1,t,\zeta-1)-\mathsf{EVSS}$, by constructing a bivariate polynomial $\mathbf{S}_n(x,y)$ such that $\mathbf{Q}_n(z) \defeq \mathbf{S}_n(0,z)=\mathbf{F}(\alpha_n)+\mathbf{R}_{1,n}x^{\zeta}+\mathbf{R}_{2,n}x^{\zeta+1}+\dots+\mathbf{R}_{t,n}x^{\zeta+t-1}$, for some $t , \zeta \in \mathbb{N}$. Note that the coefficient of $x^{\ell}$ is 0 for all $\ell \in [\zeta-1]$.

In a \emph{semi-honest} setting, it is not complicated to reach this goal. Because there exists no malicious node and each node $n$ can make polynomial $\mathbf{Q}_n(x)\defeq\mathbf{F}(\alpha_n)+\mathbf{R}_{1,n}x^{\zeta}+\mathbf{R}_{2,n}x^{\zeta+1}+\dots+\mathbf{R}_{t,n}x^{\zeta+t-1}$, and sends $\mathbf{Q}_n(\alpha_{n'})$ to node $n'$. However, in our scenario, there exist some malicious nodes and we need to force them to share the correct value in a correct polynomial form, otherwise, it will be eliminated. To detect and correct any malicious behavior in the process of sharing, we propose the following procedure. We want to verify that the constant term of $\mathbf{Q}_n(x)$ is the correct share, i.e., sampled from a degree $p$ polynomial, and they are shared via a polynomial of degree $\zeta+t-1$ where the coefficient of $x^l$ is equal to zero for $l \in [\zeta-1]$.

In addition this algorithm satisfies some privacy constraints. At this point we will not talk about those properties. Later when we use this algorithm as a subroutine in the main procedure, we will show that it does not violate the privacy constraints that we are looking for.

Each worker $n$ can use $(\mathbf{F}(\alpha_n),1,t,\zeta-1)-\mathsf{EVSS}$ to share $\mathbf{F}(\alpha_n)$. Using $\mathsf{EVSS}$ ensures other nodes that the degree of the polynomial is equal to $\zeta+t-1$. Now we aim to satisfy two additional conditions for each node $n$ which $\mathsf{EVSS}$ can not provide it alone. Each node $n$ need to verify that:

\begin{enumerate}
	\item  The constant term of $\mathbf{S}_{n'}(0,z)$ is exactly $\mathbf{F}(\alpha_{n'})$, for $n' \in [N]$.
	\item  In $\mathbf{S}_{n'}(0,z)=\mathbf{Q}_{n'}(z)$ the coefficient of $x^j$ is zero, for $j \in [\zeta-1]$.
\end{enumerate}
Otherwise, it will announce node $n'$ as an adversary node.

Let $G \in \mathbb{F}^{(t+1)\times N}$ be the generator matrix of polynomial $\mathbf{F}(x)=\displaystyle\sum_{i=0}^{p}\eta_i x^i$, i.e.,
\begin{align*}
\mathbf{G} \defeq
\begin{bmatrix}
\alpha_{1}^0 &\alpha_{2}^0 & \alpha_{3}^0 &\dots & \alpha_{N}^0\\
\alpha_{1}^{1} &\alpha_{2}^{1} & \alpha_{3}^{1} &\dots & \alpha_{N}^{1}\\
\alpha_{1}^{2} &\alpha_{2}^{2} & \alpha_{3}^{2}&\dots & \alpha_{N}^{2}\\
\vdots & \vdots & \vdots & \ddots & \vdots \\
\alpha_{1}^{p} &\alpha_{2}^{p} & \alpha_{3}^{p} &\dots & \alpha_{N}^{p}
\end{bmatrix}.
\end{align*}

One can see that, $\mathbf{G}$ is a generator for a Reed-Solomon code of length $N$, dimension $p+1$, with the  minimum distance $d=N-p$. For this code, there exists a \emph{parity-check matrix} $\mathbf{H} \in \mathbb{F}^{(d-1) \times N}$ of rank $d-1$ which is determined by $\alpha_1,\alpha_2,\dots,\alpha_{N}$ such that $\mathbf{G}\mathbf{H}^T =\mathbf{0}^{(p+1)\times (d-1)}$. Let us define $\mathbf{\beta} \defeq [\mathbf{F}(\alpha_1),\mathbf{F}(\alpha_2),\dots,\mathbf{F}(\alpha_N)]$. One can see that $\mathbf{\beta}\mathbf{H}^T=[\eta_0,\eta_1,\dots,\eta_p]\mathbf{G}\mathbf{H}^T=\mathbf{0}^{(t+1) \times (d-1)}$. For every error vector $\mathbf{e} \in \{0,1\}^N$, we have $(\mathbf{\beta} +\mathbf{e})\mathbf{H}=\mathbf{e}\mathbf{H}$. If Hamming distance of $\mathbf{e}$ is at most $\frac{d}{2}$ from $\Vec{\mathbf{0}}$, it is possible to detect and correct the errors. Since there exist at most $t$ malicious adversaries, we need to have $\frac{N-p}{2}=\frac{d}{2}>t$. In our scheme, the maximum value of $p$ is $m+t-1$. So it is enough to have $N \geq 3t+m$. By using \emph{multiplication of shares with matrix $\mathbf{H}^T$} as it was explained in the previous subsection, we can compute the product of $\mathbf{\beta}$ and $\mathbf{H}$ which ideally is equal to $\Vec{\mathbf{0}}$.
To do this, we ask each node $n$ to re-share its share using a polynomial in the form of $\mathbf{Q}_n(z)=\mathbf{F}(\alpha_n)+\mathbf{R}_{1,n}x^{\zeta}+\mathbf{R}_{2,n}x^{\zeta+1}+\dots+\mathbf{R}_{t,n}x^{\zeta+t-1}$ using $(\mathbf{F}(\alpha_n),1,t,\zeta-1)-\mathsf{EVSS}$. Therefore, similar to the previous subsection, each node $n'$ has access to $[\mathbf{Q}_1(\alpha_{n'}),\mathbf{Q}_2(\alpha_{n'}),\dots,\mathbf{Q}_N(\alpha_{n'})]$. Then, it computes $[\mathbf{Y}_{n',1},\mathbf{Y}_{n',2},\dots,\mathbf{Y}_{n',d-1}]\defeq[\mathbf{Q}_1(\alpha_{n'}),\mathbf{Q}_2(\alpha_{n'}),\dots,\mathbf{Q}_N(\alpha_{n'})]\mathbf{H}^T$ and sends $[\tilde{\mathbf{Y}}_{n',1},\tilde{\mathbf{Y}}_{n',2},\dots,\tilde{\mathbf{Y}}_{n',d-1}]$ to all other nodes. Similar to the previous subsection, by Reed-Solomon decoding, each node can derive $\mathbf{Y}(x)=[\mathbf{Y}_1(x),\mathbf{Y}_2(x),\dots$ $,\mathbf{Y}_m(x)]\defeq [Q_1(x),Q_2(x),\dots,Q_N(x)]\mathbf{H}^T$. So it can compute $[Q_1(0),Q_2(0),\dots,Q_N(0)]\mathbf{H}$ which is ideally equal to $\mathbf{\beta}\mathbf{H}^T=\Vec{\mathbf{0}}$. If it is not equal to $\Vec{\mathbf{0}}$, then honest nodes can identify malicious workers.

Up to here, we are sure that the correct value is re-shared. Now the turn is to verify that the polynomial $\mathbf{Q}_n(z)$ has the valid form, meaning  the coefficient of $x^j$ is zero, for $j \in [\zeta-1]$.
For each honest node $n$, generator matrix $\mathbf{\tilde{G}}$ of polynomial $\mathbf{Q}_n(z)$ is as follows.

\begin{align}
\label{generatorsAni}
\mathbf{\tilde{G}} \defeq
\begin{bmatrix}
\alpha_{1}^0 &\alpha_{2}^0 & \alpha_{3}^0 &\dots & \alpha_{N}^0\\
\alpha_{1}^{\zeta} &\alpha_{2}^{\zeta} & \alpha_{3}^{\zeta} &\dots & \alpha_{N}^{\zeta}\\
\alpha_{1}^{\zeta+1} &\alpha_{2}^{\zeta+1} & \alpha_{3}^{\zeta+1}&\dots & \alpha_{N}^{\zeta+1}\\
\vdots & \vdots & \vdots & \ddots & \vdots \\
\alpha_{1}^{\zeta+t-1} &\alpha_{2}^{\zeta+t-1} & \alpha_{3}^{\zeta+t-1} &\dots & \alpha_{N}^{\zeta+t-1}
\end{bmatrix}
\end{align}

There exists a parity-check matrix $\mathbf{\tilde{H}}$ for this generator matrix. Since each node $n'$ has a private value $\mathbf{Q}_n(\alpha_{n'})$ and has access to $\mathbf{\tilde{H}}$, they can invoke \emph{multiplication of shares $[\mathbf{Q}_n(\alpha_{1}),\mathbf{Q}_n(\alpha_{2}),\dots,\mathbf{Q}_n(\alpha_{N})]$ by matrix $\mathbf{\tilde{H}}^T$}. If the result is $\Vec{\mathbf{0}}$, then the form of $\mathbf{Q}_n(z)$ can be verified, otherwise, it means that the generator matrix does not follow the format of \eqref{generatorsAni}.
Later we will use this subroutine in the process of calculating an arbitrary polynomial. We show that when we use this algorithm, it does not violate the privacy constraints, stated in the problem formulation, in Section \ref{sec:Problem Setting}.

\vspace{0.4 in}
\begin{Algorithm}{Subshare of shares}
\label{protocolsubshare}
\textit{Initializing.}
Let $\alpha_1,\alpha_2,\dots,\alpha_{N} \in \mathbb{F}$ be $n$ distinct non-zero elements of $\mathbb{F}$ known to  all nodes. In addition $p \in \mathbb{N}$ is an integers known to every one. Each node $n$ holds $\mathbf{F}(\alpha_n)$, where $\mathbf{F}(x)$ is a polynomial of degree $p$. Let $\mathbf{G}$ be a generator matrix where $g_{i,j}=\alpha_i^j$, and $\mathbf{H}\in \mathbb{F}^{2k \times N}$ be the corresponding parity-check matrix. Each node also knows the value of $\zeta$ and $\deg(\mathbf{F}(x))$.

\begin{enumerate}
	\item 
	Each node $n$ constructs a polynomial $\mathbf{Q}_n(x)\defeq\mathbf{F}(\alpha_n)+\mathbf{R}_{1,n}x^{\zeta}+\mathbf{R}_{2,n}x^{\zeta+1}+\dots+\mathbf{R}_{t,n}x^{\zeta+t-1}$, where $\mathbf{R}_{i,n}, i \in [t]$, are chosen independently and uniformly at random from the field.
	\item
	
	Each node $n$ invokes Algorithm \ref{protocolMat}, using $\mathbf{Q}_n(x)$  and $\mathbf{H}$ as inputs to the algorithm. Each node $n$ recovers $\mathbf{Q}_1(\alpha_n),\mathbf{Q}_2(\alpha_n),\dots,\mathbf{Q}_N(\alpha_n)$ and $[\mathbf{Q}_1(0),\mathbf{Q}_2(0),\dots,\mathbf{Q}_N(0)]\mathbf{H}$ which is equal to $\mathbf{e}\mathbf{H}$, where $\mathbf{e}$ is the error vector (error happens if $\mathbf{Q}_{n'}(0) \neq \mathbf{F}(\alpha_{n'})$, for some $n' \in [N]$). Thus, by running Reed-Solomon decoding procedure, it can obtain error vector $\mathbf{e}$. 
\end{enumerate}
\end{Algorithm}

\vspace{0.4 in}

\begin{Algorithm}{Subshare of shares in the form of $(\mathbf{F}(\alpha_n),1,t,\zeta-1)-\mathsf{EVSS}$}
\label{verifyprotocolsubshare}
\textit{Initializing.}
Let $\alpha_1,\alpha_2,\dots,\alpha_{N} \in \mathbb{F}$ be $n$ distinct non-zero elements of $\mathbb{F}$ known to  all nodes. Each node $n$ has $\mathbf{Q}_1(\alpha_n), \mathbf{Q}_2(\alpha_n),\dots,\mathbf{Q}_N(\alpha_n)$ from Algorithm \ref{protocolsubshare}. In addition $\zeta \in \mathbb{N}$ is an integers known to every one. Let $\tilde{\mathbf{G}}$ be a generator matrix of polynomial $\mathbf{Q}_n(x)$, and $\tilde{\mathbf{H}}$ be the corresponding parity-check matrix.

For each $n$, all nodes do the following steps:

\begin{enumerate}
	\item 
	Each node $n'$ constructs a polynomial $\mathbf{T}_{n'}(x)\defeq\mathbf{Q}_n(\alpha_{n'})+\mathbf{R'}_{1,n'}x^{1}+\mathbf{R'}_{2,n'}x^{2}+\dots+\mathbf{R'}_{t,n'}x^{t}$, where $\mathbf{R'}_{i,n'}, i \in [t]$, are chosen independently and uniformly at random from the field.
	\item
	
	Each node $n'$ invokes Algorithm \ref{protocolMat}, using $\mathbf{T}_n(x)$  and $\tilde{\mathbf{H}}$ as inputs to the algorithm. Each node $n'$ recovers $\mathbf{T}_1(\alpha_{n'}),\mathbf{T}_2(\alpha_{n'}),\dots,\mathbf{T}_N(\alpha_{n'})$ and computes $[\mathbf{T}_1(0),\mathbf{T}_2(0),\dots,\mathbf{T}_N(0)]\mathbf{H}$ which is equal to $\vec{\mathbf{0}}$, otherwise, node $n$ is eliminated from the remaining part of the algorithm.
\end{enumerate}

\end{Algorithm}

\newpage

\subsection{Evaluating a shared polynomial}
\label{evaluating}

 To be able to verify the multiplication of shares in Section \ref{with}, nodes need to investigate \texttt{complaints}. To do this, nodes need to be able to evaluate shared polynomials at any predetermined point $\alpha$. Consider a polynomial $\mathbf{F}(x)$ of degree $p$. Each node $n$ has access to $\mathbf{F}(\alpha_n)$ as its private share. All nodes aim to compute $\mathbf{F}(\alpha)$ in collaboration with each other for an arbitrary $\alpha$ without revealing anything beyond that, about their secrets. In this system, at most $t$ of the nodes are malicious. Honest nodes want to identify any malicious behavior and omit identified malicious nodes from the remaining stages of computation. We will show that if the total number of nodes is equal or grater than $\max(3t,t+p+1)$, we can reach this goal.
In this functionality we use nodes which are not already determined as a malicious node. Without loss of generality, we assume that nodes  $1,2,\dots,N'$ are not identified as malicious so far.

First, by invoking \emph{Subshare of $\mathbf{F}(\alpha_n)$ in the form of} $(\mathbf{F}(\alpha_n),1,t,0)-\mathsf{EVSS}$, each honest node $n$ subshares its private input $\mathbf{F}(\alpha_n)$ using $\mathbf{Q}_n(x)=\mathbf{F}(\alpha_n)+\mathbf{R}_{1,n}x^{\zeta}+\mathbf{R}_{2,n}x^{\zeta+1}+\dots+\mathbf{R}_{t,n}x^{\zeta+t-1}$, where $\mathbf{R}_{1,n},\mathbf{R}_{2,n},\dots,\mathbf{R}_{t,n}$ are chosen independently and uniformly at random, and sends $\mathbf{Q}_n(\alpha_{n'})$ to node $n'$, for some $t , \zeta \in \mathbb{N}$ and $n,n' \in [N']$. Hence each node $n'$ has access to $\mathbf{Q}_1(\alpha_{n'}),\mathbf{Q}_2(\alpha_{n'}),\dots,\mathbf{Q}_{N'}(\alpha_{n'})$. By Theorem \ref{combination lagrange}, if $N'>\text{deg}(\mathbf{F}(x))=p$, there exist some $\lambda_1,\lambda_2,\dots,\lambda_{N'}$ such that $\mathbf{F}(\alpha)=\displaystyle\sum_{i=1}^{N'}\lambda_i \mathbf{F}(\alpha_{i})$. Each node $n$ computes $\mathbf{Q}(x)=\displaystyle\sum_{i=1}^{N'}\lambda_i \mathbf{Q}_i(x)$ at point $\alpha_n$, i.e., $\mathbf{Q}(\alpha_{n})=\displaystyle\sum_{i=1}^{N'}\lambda_i \mathbf{Q}_i(\alpha_{n})$, and then sends it to each node $n' \in [N']$. Each node $n'$ receives $\tilde{\mathbf{Q}}(\alpha_{1}),\tilde{\mathbf{Q}}(\alpha_{2}),\dots,\tilde{\mathbf{Q}}(\alpha_{N'})$. If node $n$ is honest, $\tilde{\mathbf{Q}}(\alpha_{n})$ is equal with $\mathbf{Q}(\alpha_{n})$. Note that if a node is malicious, it can send different values to different nodes. One can see that $[\tilde{\mathbf{Q}}(\alpha_{1}),\tilde{\mathbf{Q}}(\alpha_{2}),\dots,\tilde{\mathbf{Q}}(\alpha_{N'})]$ is a Reed-Solomon code of distance $N'-t$ which can correct up to $\frac{N'-t}{2}$ errors. Since there exist at most $t$ malicious nodes, we need to $\frac{N'-t}{2}>t$ on $N' > 3t$. Thus, by using Reed-Solomon decoding, each node can reconstruct $\mathbf{Q}(x)$, where $\mathbf{Q}(0)=\mathbf{F}(\alpha)$. 

\vskip .5in

\begin{Algorithm}{Evaluating a shared polynomial}
\label{protocolEval}
\textit{Initializing.} Each node $n$ holds a value $\mathbf{F}(\alpha_n)$, where $\mathbf{F}(x)$ is a polynomial of degree $p$. $N$ non-zero distinct elements $\alpha_1,\alpha_2,\dots,\alpha_{N} \in \mathbb{F}$ are known to all nodes.  For simplicity assume that nodes $\{1,2,\dots,N'\}$ are not identified as malicious so far.

\textit{ The protocol:}
\begin{enumerate}
	\item 
	Each honest node $n$ subshares its private input $\mathbf{F}(\alpha_n)$ by invoking subshare of shares algorithm and sends $\mathbf{Q}_n(\alpha_{n'})$ to node $n'$ where $\mathbf{Q}_n(0)=\mathbf{F}(\alpha_n)$. Construction of $\mathbf{Q}_n(x)$ is explaind in previous subsection.
	\item
	There exist some $\lambda_1,\lambda_2,\dots,\lambda_{N'} \in \mathbb{F}$ such that $\mathbf{F}(\alpha)=\displaystyle\sum_{i=1}^{N'}\lambda_i \mathbf{F}(\alpha_{i})$. Each node $n$ computes $\mathbf{Q}(\alpha_{n})=\displaystyle\sum_{i=1}^{i=N'}\lambda_i \mathbf{Q}_i(\alpha_{n})$ and broadcasts it to all other nodes.
	\item
	Each node $n'$ receives $\tilde{\mathbf{Q}}(\alpha_{1}),\tilde{\mathbf{Q}}(\alpha_{2}),\dots,\tilde{\mathbf{Q}}(\alpha_{N'})$. Missing values are replaced by 0. Each party runs Reed-Solomon decoding on possibly corrupted codewords $[\tilde{\mathbf{Q}}(\alpha_{1}),\tilde{\mathbf{Q}}(\alpha_{2}),\dots,\tilde{\mathbf{Q}}(\alpha_{N'})]$ and corrects the possible errors and reconstructs $[\mathbf{Q}(\alpha_{1}),\mathbf{Q}(\alpha_{2}),\dots,$ $\mathbf{Q}(\alpha_{N'})]$. Then this node reconstructs $\mathbf{Q}(x)$, where $\mathbf{Q}(0)=\mathbf{F}(\alpha)$. 
	
\end{enumerate}

\end{Algorithm}

\newpage

\section{multiplication of two matrices in presence of adversaries}
\label{with}

Assume that two matrices $\mathbf{A},\mathbf{B} \in \mathbb{F}^{z \times z}$ are partitioned as $\mathbf{A}=[\mathbf{A}_0|\mathbf{A}_1|\dots|\mathbf{A}_{m-1}]$ and $\mathbf{B}=[\mathbf{B}_0|\mathbf{B}_1|\dots|\mathbf{B}_{m-1}]$ where $\mathbf{A}_i,\mathbf{B}_i \in \mathbb{F}^{z \times \frac{z}{m}}$, for some integer $m, m|z$. We want to multiply $\mathbf{A}\text{ and }\mathbf{B}^T$ distributively over the workers. In the end, each worker needs to have $(m,t,0,\mathsf{d})$-share of $\mathbf{C}=\mathbf{A}\mathbf{B}^T$. Recall that we have $N$ workers, and up to $t$ of them are malicious. We will show that at the presence of these malicious behavior, we still can derive the correct result. In this section we follow the same protocol as it was demonstrated in Section \ref{without} with some modifications, using the subroutines explained in Section \ref{preliminaries} to detect and correct malicious behaviour of the workers.

\textbf{1- Sharing phase:} This phase is similar to the sharing phase of Section \ref{without}.

\textbf{2- Subsharing phase:} Similar to Section \ref{without}, each honest worker $n$ wants to send the value of polynomials $\mathbf{A}_n(x)$ and $\mathbf{B}_n^{(j)}(x)$, $j=0, \ldots, m-1$,  at point $\alpha_{n'}$ to worker ${n'}$. But if worker $n$ is malicious, it can send inconsistence shares to the other workers to mislead the entire system and possibly makes the final result incorrect. To prevent this we can use algorithm of subshare of shares. If $N \geq 3t+m$ each honest worker $n$ can subshare $[\Ab]_{n},  [\Bb]_{n,0}, \ldots [\Bb]_{n,m-1}$
in the form of $([\Ab]_{n},1,t,m-1)-\mathsf{EVSS},([\Bb]_{n,0},1,t,m-1)-\mathsf{EVSS},([\Bb]_{n,1},1,t,m-2)-\mathsf{EVSS},\dots,$ $([\Bb]_{n,m-1},1,t,0)-\mathsf{EVSS}$, respectively. Subshare of shares algorithm makes nodes to be sure that they have the correct values of  $(1,t,m-1,\mathsf{d})-$share of $[\Ab]_{n}$ and $(1,t,m-1-i,\mathsf{d})-$share of $[\Bb]_{n,i}, i \in \{0,1,\dots,m-1\}$.

Let us denote the bivariate polynomial which is used in the above $\mathsf{EVSS}$ invocations by $\mathbf{S}^{(0)}_n(x,y),$ $\mathbf{S}^{(1)}_n(x,y), \dots,\mathbf{S}^{(m)}_{n}(x,y)$, respectively. By following $\mathsf{EVSS}$ algorithm explained in Subsection \ref{preliminaries}, we have 
\begin{align}
\mathbf{A}_n(y) & =\mathbf{S}^{(n)}_0(0,y),\\
\mathbf{B}_n^{(j)}(y)  & =\mathbf{S}^{(n)}_{j+1}(0,y), \quad   j \in \{0,1,\dots,m-1\}. 
\end{align}
By using Subroutine \ref{subshare}, workers can be sure that worker $n$ uses $m+1$ polynomials $\mathbf{A}_n(x)$ and $\mathbf{B}_n^{(j)}(x)$ for $j \in \{0,1,\dots,m-1\}$ with degree $m+t-1$ and $m+t-j-1$, respectively, where the coefficients of $x^\ell$ in $\mathbf{A}_n(x)$ is zero for $\ell \in [m-1]$, and the coefficient of $x^{\ell'}$ in $\mathbf{B}_n^{(j)}(x)$ is equal to zero for $\ell' \in [m-j-1]$.
Let us define $\mathbf{B}_n(x)$ as

\begin{align}
\mathbf{B}_n(x)\defeq\displaystyle \sum_{k=0}^{m-1}\mathbf{B}_n^{(k)}(x)x^k=\mathbf{B}_n^{(0)}(x)+\mathbf{B}_n^{(1)}(x)x^1+\dots+\mathbf{B}_n^{(m-1)}(x)x^{m-1}.
\end{align}

Each workers $n'$ can easily calculate $\mathbf{B}_n(\alpha_{n'})$, for all ${n'} \in [N]$. Note that, we partition $\mathbf{B}_n(x)$ into $\mathbf{B}_n^{(0)}(x),$ $\mathbf{B}_n^{(1)}(x),\dots,$ $\mathbf{B}_n^{(m-1)}(x)$, because we cannot verify the coefficients of $x^0,x^1,\dots,x^{m-1}$, simultaneously, in $\mathbf{B}_n(x)$. One can see that, the coefficients of $x^0,x^1,\dots,x^{m-1}$ in $\mathbf{B}_n(x)$, are equal to the coefficient of $x^0$ in polynomials $\mathbf{B}_n^{(0)}(x),\mathbf{B}_n^{(1)}(x),\dots,$ $\mathbf{B}_n^{(m-1)}(x)$, respectively, which can be verified by using Subroutine \ref{subshare}.

\textbf{3- Constructing the product of subshares:} 
Similar to the previous phase, existence of malicious workers are the main concern  in this phase. Up to now, there is a subset of workers, that each of them has a verified version of subshares of shares of the others. The rest of the workers have been eliminated from the process, because they have been identified as adversaries.  Recall that each worker $n$ needs to have a share of $\mathbf{C}=\mathbf{A}\mathbf{B}^T$ (i.e., $\mathbf{F}^{(\mathsf{d})}_{\mathbf{A}\mathbf{B}^T,m,t,0}(\alpha_n)$, for all $n \in [N]$).
As it was shown in Section \ref{without}, there exist $m+t-1$ polynomials $\mathbf{O}_0^{(n)}(x),\mathbf{O}_1^{(n)}(x),\dots,\mathbf{O}_{m+t-2}^{(n)}(x)$ of degree $t$ such that

\begin{align}
\label{witheq4}
\mathbf{C}_n(x)\defeq \mathbf{F}^{(\mathsf{d})}_{[\mathbf{A}]_n[\mathbf{B}]_n^T,m,t,0}(x)
=
\mathbf{A}_n(x)\mathbf{B}_n(x)
-\displaystyle
\sum_{\ell=0}^{m+t-2}x^{m+\ell}
\mathbf{O}^{(n)}_{\ell}(x).
\end{align}

After constructing polynomials $\mathbf{O}_0^{(n)}(x),\mathbf{O}_1^{(n)}(x),\dots,\mathbf{O}_{m+t-2}^{(n)}(x)$, worker $n$ shares them using $\mathsf{EVSS}$. Then, it shares $\mathbf{C}_n(x)$ directly to allow each node to verify the sharing process of $\mathbf{C}_n(x)$.

From \eqref{eq4}, one can see that the coefficient of $x^j$ in $\mathbf{C}_n(x)$ is $\{[\mathbf{A}]_n[\mathbf{B}]_n^T\}_j$ for all non-negative $j<m$. This is due to the fact that each $\mathbf{O}^{(n)}_{\ell}(x)$ is multiplied by $x^{m+\ell}$ where $\ell \geq m$, thus these do not affect the coefficient of $x^j$, where $j<m$. 
Construction of each $\mathbf{O}_{\ell}^{(n)}(x)$ explained in Section \ref{with} in details.
Each honest workers $n' $ directly verifies that, if \eqref{witheq4} does not hold, then it sends \texttt{complaint} message and other workers verify this by invoking \emph{evaluating a shared polynomial} algorithm, with the constraint $N>3t+m$. In other words, if node $n'$ sends a $\mathsf{Complaint}$ message, all nodes in collaboration with each other can compute $\mathbf{A}_n(\alpha_{n'}),\mathbf{B}_n(\alpha_{n'}),\mathbf{O}_0(\alpha_{n'}),\dots,\mathbf{O}_{m+t-2}(\alpha_{n'})$ and $\mathbf{C}_n(\alpha_{n'})$. Then they can examine if \eqref{witheq4} holds at point $\alpha_{n'}$ or not. On top of that they can identify the malicious nodes. If there is no $\mathsf{complaint}$, each node $n'$ accepts $\mathbf{C}_n(\alpha_{n'})$ as the correct share of $[\mathbf{A}]_n[\mathbf{B}]_n^T$.
Assume that the dealer (here node $n$) is malicious and wants to cheat. The algorithm of subshare of shares, prevents cheating in sharing of $\mathbf{A}_n(x) \text{ and }\mathbf{B}_n(x)$. Also using EVSS makes us sure that the degree of polynomials $\mathbf{C}_n(x)$ and $\mathbf{O}^{(n)}_i(x)$ are $m+t-1$ and $t$, respectively, for all $i \in \{0,1,\dots,m+t-2\}$.
So if there exists any adversarial behaviour, the degree of $\tilde{\mathbf{C}}_n(x)\defeq\mathbf{A}_n(x)\mathbf{B}_n(x)-\displaystyle\sum_{\ell=0}^{m+t-2}x^{m+\ell}\mathbf{O}^{(n)}_{\ell}(x)$ is at most $2m+2t-2$. Assume that $\tilde{\mathbf{C}}_n(x) \neq \mathbf{C}_n(x)$. Since deg($\mathbf{C}_n(x))=m+t-1$ and deg($\tilde{\mathbf{C}}_n(x)$) is at most $2m+2t-2$, $\tilde{\mathbf{C}}_n(x)$ can have at most $2t+2m-2$ common points with $\mathbf{C}_n(x)$. So if at least $2m+2t-1$ honest nodes verify that $\tilde{\mathbf{C}}_n(x)$ is consistent with $\mathbf{C}_n(x)$, it means that $\tilde{\mathbf{C}}_n(x)=\mathbf{C}_n(x)$. Thus, if we have  $N \geq 3t+2m-1$, honest nodes can derive their shares, correctly.

\textbf{4- Constructing shares of product:} This phase is similar to the constructing shares of product matrix of Section \ref{without}.

\section{Transforming Shares}
\label{transformingshares}
In this section, we explain how to change the  parameters of the shares. This is needed to be able to handle arbitrary polynomials, as we will see later.

\subsection{Direct to Reverse Polynomial Sharing}
\label{directtoreversepolynomial}
Assume that matrix $\Ab=[\Ab_0,\Ab_1,\dots,\Ab_{m-1}] \in \mathbb{F}^{k \times k}$ is shared among $N$ nodes directly, i.e., each node $n$ has received $\mathbf{F}^{(\mathsf{d})}_{\Ab,m,t,0}(\alpha_n)=\Ab_0+\Ab_1\alpha_n+\Ab_2\alpha_n^2+\dots+\Ab_{m-1}\alpha_n^{m-1}+\mathbf{R}_1\alpha_n^m+\dots+\mathbf{R}_t\alpha_n^{m+t-1}$, where $\mathbf{R}_i$ are chosen uniformly and independently from $\mathbb{F}^{k \times \frac{k}{m}}$. All nodes aim to obtain shares of matrix $\Ab$ in reverse polynomial sharing format, i.e., each node $n$ needs to have $\mathbf{F}^{(\mathsf{r})}_{\Ab,m,t,0}(\alpha_n)$. To this end, each node $n$ is enough to have the value of $m$ polynomials $\Ab^{(0)}(x),\Ab^{(1)}(x),\dots,\Ab^{(m-1)}(x)$ at point $\alpha_n$ where,

\begin{align}
&\Ab^{(0)}(x)\defeq \Ab_0+\mathbf{R}_1^{(0)}x+\mathbf{R}_2^{(0)}x^2+\dots+\mathbf{R}_{t}^{(0)}x^t=\mathbf{F}^{(\mathsf{d})}_{\Ab_0,1,t,0}\\
&\Ab^{(1)}(x)\defeq
\Ab_1+\mathbf{R}_1^{(1)}x^2+\mathbf{R}_2^{(1)}x^3+\dots+\mathbf{R}_{t}^{(1)}x^{t+1}=\mathbf{F}^{(\mathsf{d})}_{\Ab_1,1,t,1}\\
&\Ab^{(2)}(x)\defeq
\Ab_2+\mathbf{R}_1^{(2)}x^3+\mathbf{R}_2^{(2)}x^4+\dots+\mathbf{R}_{t}^{(2)}x^{t+2}=\mathbf{F}^{(\mathsf{d})}_{\Ab_2,1,t,2}\\
&\vdots\\
&\Ab^{(m-1)}(x)\defeq
\Ab_{m-1}+\mathbf{R}_1^{(m-1)}x^m+\mathbf{R}_2^{(m-1)}x^{m+1}+\dots+\mathbf{R}_{t}^{(m-1)}x^{m+t-1}=\mathbf{F}^{(\mathsf{d})}_{\Ab_{m-1},1,t,m-1},
\end{align}
where $\mathbf{R}^{(i)}_j$ are chosen uniformly and independently at random from $\mathbb{F}^{k \times \frac{k}{m}}$, for all $i \in \{0,1,\dots,m-1\} \text{ and } j \in [t]$. Then, node $n$ computes the value of $\Ab^{(m-1)}(x)+x\Ab^{(m-2)}(x)+x^2\Ab^{(m-3)}(x)+\dots+x^{m-1}\Ab^{(0)}(x)$ at point $\alpha_n$, which is equal to  $\mathbf{F}^{(\mathsf{r})}_{\Ab,m,t,0}(\alpha_n)$, as shown here:

\begin{align}
\begin{split}
\label{formulacombination}
\Ab^{(m-1)}(x)+&x\Ab^{(m-2)}(x)+x^2\Ab^{(m-3)}(x)+\dots+x^{m-1}\Ab^{(0)}(x)\\= &\Ab_{m-1}+\Ab_{m-2}x+\Ab_{m-3}x^2+\dots+\Ab_{0}x^{m-1}+(\displaystyle\sum_{i=0}^{m-1}\mathbf{R}^{(i)}_1)x^m+\dots+(\displaystyle\sum_{i=0}^{m-1}\mathbf{R}^{(i)}_{t})x^{m+t-1}\\=
&\Ab_{m-1}+\Ab_{m-2}x+\Ab_{m-3}x^2+\dots+\Ab_{0}x^{m-1}+ \mathbf{Z}_1x^m+\mathbf{Z}_2x^{m+1}+\dots+\mathbf{Z}_tx^{m+t-1}\\=&\mathbf{F}^{(\mathsf{r})}_{\Ab,m,t,0}(x),    
\end{split}
\end{align}
where $\mathbf{Z}_j \defeq \displaystyle\sum_{i=0}^{m-1}\mathbf{R}^{(i)}_j$.

To do this, we follow a 2-phase scheme.

\textbf{1- Subsharing Phase:} Each node $n$ has $\mathbf{F}^{(\mathsf{d})}_{\Ab,m,t,0}(\alpha_n)$ as its own share, and sub-shares it, in the form of $(\mathbf{F}^{(\mathsf{d})}_{\Ab,m,t,0}(\alpha_n),1,t,k-1)-\mathsf{EVSS}$, according to Algorithm \ref{subshare}. Then each node $n'$ has access to the value of $\mathbf{F}^{(\mathsf{d})}_{\mathbf{F}^{(\mathsf{d})}_{\Ab,m,t,0}(\alpha_n),1,t,k-1}(x) = \mathbf{F}^{(\mathsf{d})}_{\Ab,m,t,0}(\alpha_n)+\mathbf{R}_1x^k+\mathbf{R}_2x^{k+1}+\dots+\mathbf{R}_tx^{k+t-1}$ at point $\alpha_{n'}$, for all $n,n' \in [N]$. We will see that if node $n'$ combines $\mathbf{F}^{(\mathsf{d})}_{\mathbf{F}^{(\mathsf{d})}_{\Ab,m,t,0}(\alpha_1),1,t,k-1}(\alpha_{n'}),
$ $\mathbf{F}^{(\mathsf{d})}_{\mathbf{F}^{(\mathsf{d})}_{\Ab,m,t,0}(\alpha_2),1,t,k-1}(\alpha_{n'}),\dots,\mathbf{F}^{(\mathsf{d})}_{\mathbf{F}^{(\mathsf{d})}_{\Ab,m,t,0}(\alpha_N),1,t,k-1}(\alpha_{n'})$ with right coefficients, it can calculate $\Ab^{(k)}(\alpha_{n'})$. Each node $n$ does the similar scheme, for all $k \in \{0,1,\dots,m\}$.

\textbf{2- Constructing Shares Phase:}
For each $k \in \{0,1,\dots,m-1\}$ and each node $n'$, it has access to  $\mathbf{F}^{(\mathsf{d})}_{\mathbf{F}^{(\mathsf{d})}_{\Ab,m,t,0}(\alpha_n),1,t,k-1}(\alpha_{n'}) = \mathbf{F}^{(\mathsf{d})}_{\Ab,m,t,0}(\alpha_n)+\mathbf{R}_1\alpha_{n'}^k+\mathbf{R}_2\alpha_{n'}^{k+1}+\dots+\mathbf{R}_t\alpha_{n'}^{k+t-1}$. Since $\Ab_i$ is the coefficient of $x^i$ in $\mathbf{F}^{(\mathsf{d})}_{\Ab,m,t,0}(x)$, if $N \geq m+t$, there exist $\lambda_1,\lambda_2,\dots,\lambda_N$ such that $\displaystyle\sum_{i=1}^{N}\lambda_i\mathbf{F}^{(\mathsf{d})}_{\Ab,m,t,0}(\alpha_i)=\Ab_{i}$. Each node $n'$ computes $\displaystyle\sum_{i=1}^{N}\lambda_i\mathbf{F}^{(\mathsf{d})}_{\mathbf{F}^{(\mathsf{d})}_{\Ab,m,t,0}(\alpha_i),1,t,k-1}(\alpha_{n'})$ to derive $\mathbf{A}^{(k)}(x)=\mathbf{F}^{(\mathsf{d})}_{\Ab_i,1,t,k-1}(\alpha_{n'})$. This scheme enables node $n$ to derive $\Ab^{(0)}(\alpha_n),$ $\Ab^{(1)}(\alpha_n),\dots,\Ab^{(m-1)}(\alpha_n)$. Finally, as mentioned in \eqref{formulacombination}, it can compute $\mathbf{F}^{(\mathsf{r})}_{\Ab,m,t,0}(\alpha_n)$.

\begin{Remark}
To obtain direct polynomial shares from reverse polynomial sharing format, we can follow similar steps.
\end{Remark}

\subsection{Shares of Transposed Matrix}
\label{Sharesoftrransposedmatrix}
Assume that matrix $\Ab=[\Ab_0,\Ab_1,\dots,\Ab_{m-1}] \in \mathbb{F}^{k \times k}$ is shared among $N$ nodes. It means that each node $n$ has the value of polynomial $\mathbf{F}^{(\mathsf{d})}_{\Ab,m,t,0}(x)\defeq \Ab_0+\Ab_1x+\Ab_2x^2\dots+\Ab_{m-1}x^{m-1}+\mathbf{R}_1x^m+\mathbf{R}_2x^{m+1}+\dots+\mathbf{R}_tx^{m+t-1}$ at $x=\alpha_n$ where $\mathbf{R}_i$ are chosen uniformly and independently at random from $\mathbb{F}^{k \times\frac{k}{m}}$. The goal is that each node has a share of matrix $\Ab^T$.
Let us partition $\Ab$ as follows.

\begin{align}
\mathbf{A}&=\begin{bmatrix}
\Ab_{0, 0} & \Ab_{0, 1} & \dots &   \Ab_{0,m-1}\\
\Ab_{1, 0} & \Ab_{1, 1}& \dots &   \Ab_{1,m-1}\\
\vdots  &   \vdots  &\ddots &\vdots\\
\Ab_{m-1, 0} & \Ab_{m-1, 1}& \dots &   \Ab_{m-1,m-1}
\end{bmatrix}, \\
\end{align}
where $\Ab_{i, j}\in \mathbb{F}^{\frac{k}{m} \times \frac{k}{m}}$ for $i,j \in \{0,1,...,m-1\}$.
To share matrix $\Ab^T$, each node $n$ needs $\Ab^{(\mathsf{T})}(\alpha_n)$ where,

\begin{align}
\Ab^{(\mathsf{T})}(x)=
\begin{bmatrix}
\Ab_{0, 0}\\ \Ab_{0, 1}\\ \vdots \\ \Ab_{0,m-1}
\end{bmatrix}+
\begin{bmatrix}
\Ab_{1, 0}\\ \Ab_{1, 1}\\ \vdots \\ \Ab_{1,m-1}
\end{bmatrix}x+\dots
\begin{bmatrix}
\Ab_{m-1, 0}\\ \Ab_{m-1, 1}\\ \vdots \\ \Ab_{m-1,m-1}
\end{bmatrix}x^{m-1}+
\mathbf{R}'_1x^m+\mathbf{R}'_2x^{m+1}+\dots+\mathbf{R}'_tx^{m+t-1}.
\end{align}

To reach to our goal, each node $n$ needs to have the value of $m$ polynomials $\Ab^{(0)}(x),\Ab^{(1)}(x),\dots$ $,\Ab^{(m-1)}(x)$ at point $\alpha_n$, where

\begin{align}
&\Ab^{(m-1,0)}(x)\defeq \Ab_{m-1,0}+\mathbf{R}_1^{(m-1,0)}x+\mathbf{R}_2^{(m-1,0)}x^2+\dots+\mathbf{R}_{t}^{(m-1,0)}x^t=\mathbf{F}^{(\mathsf{d})}_{\Ab_{m-1,0},1,t,0}\\
&\Ab^{(m-2,0)}(x)\defeq
\Ab_{m-2,0}+\mathbf{R}_1^{(m-2,0)}x^2+\mathbf{R}_2^{(m-2,0)}x^3+\dots+\mathbf{R}_{t}^{(m-2,0)}x^{t+1}=\mathbf{F}^{(\mathsf{d})}_{\Ab_{m-2,0},1,t,1}\\
&\Ab^{(m-3,0)}(x)\defeq
\Ab_{m-3,0}+\mathbf{R}_1^{(m-3,0)}x^3+\mathbf{R}_2^{(m-3,0)}x^4+\dots+\mathbf{R}_{t}^{(m-3,0)}x^{t+2}=\mathbf{F}^{(\mathsf{d})}_{\Ab_{m-3,0},1,t,2}\\
&\vdots\\
&\Ab^{(0,0)}(x)\defeq
\Ab_{0,0}+\mathbf{R}_1^{(0,0)}x^m+\mathbf{R}_2^{(0,0)}x^{m+1}+\dots+\mathbf{R}_{t}^{(0,0)}x^{m+t-1}=\mathbf{F}^{(\mathsf{d})}_{\Ab_{0,0},1,t,m-1},
\end{align}
where $\mathbf{R}^{(i,l)}_j \in \mathbb{F}^{\frac{k}{m}\times\frac{k}{m}}$ are chosen uniformly and independently at random, for all $i,l \in \{0,1,\dots,m-1\} \text{ and } j \in [t]$. Then, each node $n$ computes the value of $\Ab^{(0)}(x)=\Ab^{(0,0)}(x)+x\Ab^{(1,0)}(x)+x^2\Ab^{(2,0)}(x)+\dots+x^{m-1}\Ab^{(m-1,0)}(x)$ at point $\alpha_n$ to derive $\mathbf{F}^{(\mathsf{r})}_{[\Ab_{0,0},\Ab_{1,0},\dots,\Ab_{m-1,0}],m,t,0}(\alpha_n)$, as shown here:

\begin{align}
\begin{split}
\label{formulacombinationtransposed}
\Ab^{(0)}(x)=&\Ab^{(0,0)}(x)+x\Ab^{(1,0)}(x)+x^2\Ab^{(2,0)}(x)+\dots+x^{m-1}\Ab^{(m-1,0)}(x)\\= &\Ab_{0,0}+\Ab_{1,0}x+\Ab_{2,0}x^2+\dots+\Ab_{m-1,0}x^{m-1}+(\displaystyle\sum_{i=0}^{m-1}\mathbf{R}^{(i,0)}_1)x^m+\dots+(\displaystyle\sum_{i=0}^{m-1}\mathbf{R}^{(i,0)}_{t})x^{m+t-1}\\=
&\Ab_{0,0}+\Ab_{1,0}x+\Ab_{2,0}x^2+\dots+\Ab_{m-1,0}x^{m-1}+ \mathbf{Z}_{1}^{(0)}x^m+\mathbf{Z}_{2}^{(0)}x^{m+1}+\dots+\mathbf{Z}_t^{(0)}x^{m+t-1}\\=&\mathbf{F}^{(\mathsf{r})}_{[\Ab_{0,0},\Ab_{1,0},\dots,\Ab_{m-1,0}],m,t,0}(x),
\end{split}
\end{align}
where $\mathbf{Z}_{j}^{(0)}=\displaystyle\sum_{i=0}^{m-1}\mathbf{R}^{(i,0)}_{j}$.

Clearly, $\Ab^{(0)}(\alpha_n)$ is direct polynomial share of first row of matrix $\Ab^T$. Node $n$ follows the same steps in collaboration with others, to obtain $\Ab^{(1)}(\alpha_n),\Ab^{(2)}(\alpha_n),\dots,$ and $\Ab^{(m-1)}(\alpha_n)$. Finally it can make its share equal to
$\begin{bmatrix}
\Ab^{(0)}(\alpha_n)\\
\Ab^{(1)}(\alpha_n)\\
\vdots\\
\Ab^{(m-1)}(\alpha_n)
\end{bmatrix}$.
To obtain $\Ab^{(0)}(\alpha_n)$ we can follow a 2-phase scheme.

\textbf{1- Subsharing Phase:} Each node $n$ has $\mathbf{F}^{(\mathsf{d})}_{\Ab,m,t,0}(\alpha_n)=\begin{bmatrix}
\mathbf{F}^{(\mathsf{d})}_{\Ab,m,t,0}(\alpha_n)_0\\
\mathbf{F}^{(\mathsf{d})}_{\Ab,m,t,0}(\alpha_n)_1\\
\vdots\\
\mathbf{F}^{(\mathsf{d})}_{\Ab,m,t,0}(\alpha_n)_{m-1}
\end{bmatrix}$ as its own share. To construct $\Ab^{(i,0)}(x)$, node $n$ subshares $\mathbf{F}^{(\mathsf{d})}_{\Ab,m,t,0}(\alpha_n)_i$ in the form of $(\mathbf{F}^{(\mathsf{d})}_{\Ab,m,t,0}(\alpha_n)_i,1,t,m-i-1)-\mathsf{EVSS}$, according to Algorithm \ref{subshare}. Then each node $n'$ has access to the value of $\mathbf{F}^{(\mathsf{d})}_{\mathbf{F}^{(\mathsf{d})}_{\Ab,m,t,0}(\alpha_n)_i,1,t,m-i-1}(x) = \mathbf{F}^{(\mathsf{d})}_{\Ab,m,t,0}(\alpha_n)_i+\mathbf{R}_1x^{m-i}+\mathbf{R}_2x^{m-i+1}+\dots+\mathbf{R}_tx^{m-i+t-1}$ at point $\alpha_{n'}$, for all $n,n' \in [N]$. 

We will see that if node $n'$ combines $\mathbf{F}^{(\mathsf{d})}_{\mathbf{F}^{(\mathsf{d})}_{\Ab,m,t,0}(\alpha_1)_i,1,t,m-i-1}(\alpha_{n'}),
$ $\mathbf{F}^{(\mathsf{d})}_{\mathbf{F}^{(\mathsf{d})}_{\Ab,m,t,0}(\alpha_2)_i,1,t,m-i-1}(\alpha_{n'}),\dots,$ $\mathbf{F}^{(\mathsf{d})}_{\mathbf{F}^{(\mathsf{d})}_{\Ab,m,t,0}(\alpha_N)_i,1,t,m-i-1}(\alpha_{n'})$ with right coefficients, it can derive $\Ab^{(i,0)}(\alpha_{n'})$. Each node $n$ follows the same steps to compute $\Ab^{(i,0)}(\alpha_{n})$.

\textbf{2- Constructing Shares Phase:}
For each $i \in \{0,1,\dots,m-1\}$, each node $n'$ has access to 
\begin{align}
\mathbf{F}^{(\mathsf{d})}_{\mathbf{F}^{(\mathsf{d})}_{\Ab,m,t,0}(\alpha_n)_i,1,t,m-i-1}(\alpha_{n'}) = \mathbf{F}^{(\mathsf{d})}_{\Ab,m,t,0}(\alpha_n)_i+\mathbf{R}_1\alpha_{n'}^{m-i}+\mathbf{R}_2\alpha_{n'}^{m-i+1}+\dots+\mathbf{R}_t\alpha_{n'}^{m-i+t-1}.
\end{align}
If $N \geq m+t$, there exist $\lambda_1,\lambda_2,\dots,\lambda_N$ such that $\displaystyle\sum_{j=1}^{N}\lambda_j\mathbf{F}^{(\mathsf{d})}_{\Ab,m,t,0}(\alpha_j)_i=\Ab_{i,0}$. Each node $n'$ computes $\displaystyle\sum_{j=1}^{N}\lambda_j\mathbf{F}^{(\mathsf{d})}_{\mathbf{F}^{(\mathsf{d})}_{\Ab,m,t,0}(\alpha_j)_i,1,t,m-i-1}(\alpha_{n'})$ to derive $\mathbf{A}^{(i,0)}(x)=\mathbf{F}^{(\mathsf{d})}_{\Ab_{i,0},1,t,m-i-1}(\alpha_{n'})$. This scheme enables node $n$ to achieve $\Ab^{(0,0)}(\alpha_n),\Ab^{(1,0)}(\alpha_n),\dots,\Ab^{(m-1,0)}(\alpha_n)$. Finally, as explained in \eqref{formulacombinationtransposed}, it can compute $\Ab^{(0)}(x)$.

\section{Computing an arbitrary polynomial}
\label{computinganarbitrarypolynomial}

Any polynomial of input matrices can be computed as a combination of operations like pair-wise matrix multiplication, matrix addition, matrix transposing and changing the parameters of sharing.

In Section \ref{polynomialsharing}, we show how to  share input matrices using polynomial sharing. In Section \ref{with}, we show how to multiply two matrices where the polynomial shares of those matrices are available at the nodes, and finally at the end, each node has a polynomial share of the final result. In addition, we prove that the result of the multiplication is correct. In Section \ref{transformingshares}, we show that if workers have the polynomial shares of a matrix, how nodes can communicate with each other such that each node can derive correctly the polynomial share of the transpose of that matrix without leaking any information. In Section \ref{transformingshares}, we also show that if each node has a polynomial share of a matrix, how nodes can communicate with each other to have the polynomial shares of the same matrix with different parameter correctly without leaking information.

If we have a polynomial share of two matrices with the same parameters, if each node simply adds its shares, it will have the polynomial share of the addition of those matrices. Therefore, we can apply those algorithms, one after the other, to compute the polynomial share of the final result correctly. If $N\geq 3t+2m-1$, we can manage such that each node $n$ has $\mathbf{F}^{\mathsf{(d)}}_{result,m,t,0}(\alpha_n)$ which is the polynomial share of the final result. If each of the nodes sends its share to the data collector, it can recover the final result even though $t$ of the nodes are adversary. The above arguments establish the correctness of the final result.

In Section \ref{preliminaries}, Section \ref{with} and Section \ref{transformingshares} we have not proved that the proposed algorithms and subroutines satisfy the privacy constraints. In the next section we prove the privacy constraints of the proposed scheme.

\section{proof of privacy for the proposed scheme}
\label{security}

\subsection{Overview of Privacy}

Let us assume that the goal is to compute an arbitrary function of inputs $\mathcal{X}=\{\mathbf{X}^{[1]},\mathbf{X}^{[2]},\dots,\mathbf{X}^{[\Gamma]}\}$. This computation is done in $\Delta \in \mathbb{N}$ rounds, where in each round one operation of sharing, multiplication, addition, changing the parameter of sharing, or transposing is performed, using the algorithms or subroutines explained in the paper.
Let us define $\mathcal{Y}_{\mathcal{S}}^{(\delta)}$ as the set of all messages that malicious workers receive during round $\delta$. To prove that the algorithm has no data leakage to the workers, we need to prove that
\begin{align*}
I(\mathcal{Y}_{\mathcal{S}}^{(1)},\mathcal{Y}_{\mathcal{S}}^{(2)},...,\mathcal{Y}_{\mathcal{S}}^{(\Delta)};\mathcal{X})=H(  \mathcal{Y}_{\mathcal{S}}^{(1)},\mathcal{Y}_{\mathcal{S}}^{(2)},...,\mathcal{Y}_{\mathcal{S}}^{(\Delta)})-H(  \mathcal{Y}_{\mathcal{S}}^{(1)},\mathcal{Y}_{\mathcal{S}}^{(2)},...,\mathcal{Y}_{\mathcal{S}}^{(\Delta)}|\mathcal{X})=0
\end{align*}

In other words, we need to prove that,
\begin{align}
\label{privacygeneralcase}
H(  \mathcal{Y}_{\mathcal{S}}^{(1)},\mathcal{Y}_{\mathcal{S}}^{(2)},...,\mathcal{Y}_{\mathcal{S}}^{(\Delta)})=
&H(  \mathcal{Y}_{\mathcal{S}}^{(1)},\mathcal{Y}_{\mathcal{S}}^{(2)},...,\mathcal{Y}_{\mathcal{S}}^{(\Delta)}|\mathcal{X}) \\=&
\displaystyle\sum_{\delta=1}^{\Delta}H(\mathcal{Y}_{\mathcal{S}}^{(\delta)}|\mathcal{X},  \mathcal{Y}_{\mathcal{S}}^{(1)},\mathcal{Y}_{\mathcal{S}}^{(2)},...,\mathcal{Y}_{\mathcal{S}}^{(\delta-1)}).
\end{align}

If we can prove that  $H(\mathcal{Y}_{\mathcal{S}}^{(\delta)}|\mathcal{X},  \mathcal{Y}_{\mathcal{S}}^{(1)},\mathcal{Y}_{\mathcal{S}}^{(2)},...,\mathcal{Y}_{\mathcal{S}}^{(\delta-1)})=H(\mathcal{Y}_{\mathcal{S}}^{(\delta)})$ for $\delta=1,2,\dots,\Delta$, then \eqref{privacygeneralcase} is correct. 
In what follows, we prove \eqref{privacygeneralcase} for the cases  if in round $\delta$ the system does sharing using EVSS (Subsection \ref{EVSS}), multiplication of two matrices (Section \ref{with}), transposing a matrix (Subsection \ref{Sharesoftrransposedmatrix}) and changing the parameter of sharing (Subsection \ref{directtoreversepolynomial}).

\subsection{Proof of Privacy for Sharing:}
Assume that at round $\delta$, one of the sources, e.g., source $\gamma$ shares matrix $\mathbf{X}^{[\gamma]}=[\mathbf{X}_{0}^{[\gamma]},\mathbf{X}_1^{[\gamma]},\dots,\mathbf{X}_{m-1}^{[\gamma]}]$ to be able to perform some computation on it in the next rounds. At the beginning of this round each worker $n$ has $\mathbf{F}_{X^{[\gamma]},m,t,0}^{(\mathsf{d})}(x)=\mathbf{X}_{0}^{[\gamma]}+\mathbf{X}_{1}^{[\gamma]}x+\dots+\mathbf{X}_{m-1}^{[\gamma]}x^{m-1}+R^{(\delta,\gamma,0)}x^m+R^{(\delta,\gamma,1)}x^{m+1}+\dots+R^{(\delta,\gamma,t-1)}x^{m+t-1}$ for $x=\alpha_n$.
 Let us define $\mathcal{R}^{(\delta,0)}$ as follows
\begin{align}
\mathcal{R}^{(\delta,0)} =\{\mathbf{R}^{(\delta,\gamma,\tau)}| \gamma \in [\Gamma]\text{ and } \tau \in \{ 0 , 1 , \dots, t-1 \} \}.
\end{align}

In fact, $\mathcal{R}^{(\delta,0)}$ is the set of all random matrices that source uses in this phase.

\begin{Remark}
We know that the number of malicious nodes $|\mathcal{S}|$ is at most $t$. First, in case of $|\mathcal{S}|=t$, we prove the privacy. In the second case, we assume that $|\mathcal{S} |\leq t$. Obviously, the set of all messages that malicious nodes can achieve in the second case, is a subset of the set of the data in the first case. Thus, if we prove the privacy in the first case, then it prove the privacy of the second case, automatically.
\end{Remark}

Let ${\mathcal{Y}}^{(\delta)}_\mathcal{S}$ be the set of all messages that malicious nodes $\mathcal{S}=\{s_1,s_2,\dots,s_t\}$ receive from source $\gamma$ in this step. One can see that the size of sets ${\mathcal{Y}}_{\mathcal{S}}^{(\delta)}$ and $\mathcal{R}^{(\delta,0)}$ are equal with each other which is $t \times \Gamma$. It means that the number of independent messages is equal with the number of random matrices that source $\gamma$ have been used. Thus, there is one to one corresponding between pairs $(\mathcal{X},\mathcal{R}^{(\delta,0)})$ and $(\mathcal{X},{\mathcal{Y}}_{\mathcal{S}}^{(\delta)})$. Therefore, $ H({\mathcal{Y}}_{\mathcal{S}}^{\delta}|\mathcal{X})= H(\mathcal{R}^{(\delta,0)}|\mathcal{X})$. Finally we can conclude that,
\begin{align}
H({\mathcal{Y}}_{\mathcal{S}}^{(\delta)})|\mathcal{X},\mathcal{Y}_{\mathcal{S}}^{(1)},\mathcal{Y}_{\mathcal{S}}^{(2)},...,\mathcal{Y}_{\mathcal{S}}^{(\delta-1)})&=H(\mathcal{R}^{(\delta,0)}|\mathcal{X},\mathcal{Y}_{\mathcal{S}}^{(1)},\mathcal{Y}_{\mathcal{S}}^{(2)},...,\mathcal{Y}_{\mathcal{S}}^{(\delta-1)})\\ &= H(\mathcal{R}^{(\delta,0)}|\mathcal{X})=H(\mathcal{R}^{(\delta,0)}) \overset{(a)}{\geq} H({\mathcal{Y}}_{\mathcal{S}}^{(\delta)})),
\end{align}
where $(a)$ follows from the fact that ${\mathcal{Y}}_{\mathcal{S}}^{(\delta)}$ and $\mathcal{R}^{(\delta,0)}$ have the same size, while $\mathcal{R}^{(\delta,0)}$ is selected uniformly and independently at random from the set of all matrices with the same size.

\subsection{Proof of Privacy for Multiplication}
Here we prove that if in round $\delta$ the operation is multiplication of two matrices, then
\begin{align}
H({\mathcal{Y}}_{\mathcal{S}}^{(\delta)})|\mathcal{X},\mathcal{Y}_{\mathcal{S}}^{(1)},\mathcal{Y}_{\mathcal{S}}^{(2)},...,\mathcal{Y}_{\mathcal{S}}^{(\delta-1)})= H({\mathcal{Y}}_{\mathcal{S}}^{(\delta)}).
\end{align}
 We go through phases of the algorithm for multiplication of two matrices $\mathbf{A}$, and $\mathbf{B}$.

\subsubsection{Subsharig phase}
\label{privacysubshare}

\textbf{Step 1 (EVSS)}: Node $n$ has access to $[\Ab]_n=\mathbf{F}^{(\mathsf{d})}_{\mathbf{A},m,t,0}(\alpha_n)$ and $  [\Bb]_n=
\begin{bmatrix}
&[\Bb]_{n,0}\\
&[\Bb]_{n,1}\\
&\vdots\\
&[\Bb]_{n,m-1}
\end{bmatrix}=\mathbf{F}^{(\mathsf{r})}_{\mathbf{B},m,t,0}(\alpha_n)$. This node subshares $[\Ab]_n$ and $[\Bb]_{n,j}$, for all $ j \in \{0,1,2,\dots, m-1\}$. To share $[\Ab]_n$, node $n$ uses a bivariate polynomial $\mathbf{S}_n(x,y)=\displaystyle\sum_{j=0}^{m+t-1}\displaystyle\sum_{i=0}^{m+t-1}\mathbf{S}_{i,j}x^iy^j$ such that $\mathbf{S}_{0,j}=[\Ab]_j$ for all $0 \leq j < m-1$ and $\mathbf{S}_{i,j},$ are selected independently and uniformly at random from $\mathbb{F}^{k \times \frac{k}{m}}$, for all $i \neq 0 \text{ or } m \leq j$. We partition these random coefficients into two sets. First, let $\mathcal{R}^{(\delta,\mathbf{A})}=\{\mathbf{S}_{i,j}| (i \overset{m+t}{\oplus} j \in \{ t,t+1,\dots, m+t-1\}) \text{ or } (i \geq m) \}$, where $\overset{m+t}{\oplus}$ denotes addition mod $m+t$, and the remaining coefficients are in  set $\mathcal{R}'^{(\delta,\mathbf{A})}$. Note that $|\mathcal{R}^{(\delta,\mathbf{A})}|=mt+(m+t)t=t^2+2mt$ and $|\mathcal{R}'^{(\delta,\mathbf{A})}|=m^2 -m$.
As it was mentioned in \ref{EVSS}, the number of independent equations that malicious workers receive during each EVSS invocation is equal to $t(m+t)+tm=t^2+2m$. One can see that the number of independent equations is less than the number of random matrices. We show the set of all independent equations (independent messages) which malicious nodes receive during the sharing of $\Ab$ by $\mathcal{Y}_{\mathcal{S}}^{(\delta,\mathbf{A})}$. Assume that the set $\mathcal{R}'^{(\delta,\mathbf{A})}$ is known for malicious adversaries. 
One can see that (1) $\mathcal{Y}_{\mathcal{S}}^{(\delta,\mathbf{A})}$ is a function of $  \mathcal{Y}_{\mathcal{S}}^{(1)},\mathcal{Y}_{\mathcal{S}}^{(2)},...,\mathcal{Y}_{\mathcal{S}}^{(\delta-1)}, \mathcal{R}'^{(\delta,\mathbf{A})},$ and $\mathcal{R}^{(\delta,\mathbf{A})}$. (2) Given $  \mathcal{Y}_{\mathcal{S}}^{(1)},\mathcal{Y}_{\mathcal{S}}^{(2)},...,\mathcal{Y}_{\mathcal{S}}^{(\delta-1)}, \mathcal{R}'^{(\delta,\mathbf{A})},$ there is a one to one mapping between $\mathcal{Y}_{\mathcal{S}}^{(\delta,\mathbf{A})}$ and $\mathcal{R}^{(\delta,\mathbf{A})}$. Thus,

\begin{align}
H(\mathcal{Y}_{\mathcal{S}}^{(\delta,\mathbf{A})}|\mathcal{X},  \mathcal{Y}_{\mathcal{S}}^{(1)},\mathcal{Y}_{\mathcal{S}}^{(2)},...,\mathcal{Y}_{\mathcal{S}}^{(\delta-1)})
\geq &H(\mathcal{Y}_{\mathcal{S}}^{(\delta,\mathbf{A})}|\mathcal{X},  \mathcal{Y}_{\mathcal{S}}^{(1)},\mathcal{Y}_{\mathcal{S}}^{(2)},...,\mathcal{Y}_{\mathcal{S}}^{(\delta-1)},\mathcal{R}'^{(\delta,\mathbf{A})})\\=
&H(\mathcal{R}^{(\delta,\mathbf{A})}|\mathcal{X},  \mathcal{Y}_{\mathcal{S}}^{(1)},\mathcal{Y}_{\mathcal{S}}^{(2)},...,\mathcal{Y}_{\mathcal{S}}^{(\delta-1)},\mathcal{R}'^{(\delta,\mathbf{A})})\\
=&H(\mathcal{R}^{(\delta,\mathbf{A})})
 \overset{(a)}{\geq} H(\mathcal{Y}_{\mathcal{S}}^{(\delta,\mathbf{A})}),
\end{align}
where $(a)$ is valid because $\mathcal{Y}_{\mathcal{S}}^{(\delta,\mathbf{A})}$ have the same size with $\mathcal{R}^{(\delta,\mathbf{A})}$, and $\mathcal{R}^{(\delta,\mathbf{A})}$ is selected uniformly and independently at random from the set of all matrices with the same size.
We follow the same steps for $[\Bb]_n$ and construct $\mathcal{Y}_{\mathcal{S}}^{(\delta,\mathbf{B})}, \mathcal{R}^{(\delta,\mathbf{B})}$ and $\mathcal{R}'^{(\delta,\mathbf{B})}$. Thus we have

\begin{align}
H(\mathcal{Y}_{\mathcal{S}}^{(\delta,\mathbf{B})}|\mathcal{X},  \mathcal{Y}_{\mathcal{S}}^{(1)},\mathcal{Y}_{\mathcal{S}}^{(2)},...,\mathcal{Y}_{\mathcal{S}}^{(\delta-1)},\mathcal{Y}_{\mathcal{S}}^{(\delta,\mathbf{A})})
 \geq & H(\mathcal{Y}_{\mathcal{S}}^{(\delta,\mathbf{B})}|\mathcal{X},  \mathcal{Y}_{\mathcal{S}}^{(1)},\mathcal{Y}_{\mathcal{S}}^{(2)},...,\mathcal{Y}_{\mathcal{S}}^{(\delta-1)},\mathcal{Y}_{\mathcal{S}}^{(\delta,\mathbf{A})},\mathcal{R}'^{(\delta,\mathbf{B})})\\
 = &H(\mathcal{R}^{(\delta,\mathbf{B})}|\mathcal{X},  \mathcal{Y}_{\mathcal{S}}^{(1)},\mathcal{Y}_{\mathcal{S}}^{(2)},...,\mathcal{Y}_{\mathcal{S}}^{(\delta-1)},\mathcal{Y}_{\mathcal{S}}^{(\delta,\mathbf{A})},\mathcal{R}'^{(\delta,\mathbf{B})})\\
 =&H(\mathcal{R}^{(\delta,\mathbf{B})})
  \overset{(a)}{\geq} H(\mathcal{Y}_{\mathcal{S}}^{(\delta,\mathbf{B})}).
\end{align}

\textbf{Step 2}: To be able to verify subshare of shares (Algorithm \ref{protocolsubshare}), each worker broadcasts the product of it's subshares vector and parity-check matrix $\mathbf{H}$. In other words, as mentioned in Subsection \ref{protocolMat}, each honest node $n$ sends the value of $\mathbf{Y}(x)=[\mathbf{Y}_1(x),\mathbf{Y}_2(x),\dots,\mathbf{Y}_m(x)]\defeq[Q_1(x),Q_2(x),\dots,Q_N(x)]\mathbf{H}$ at point $\alpha_n$ to all other nodes. One can see that each $\mathbf{Y}_i(x)$ has exactly $t+1$ non-zero coefficients. Each malicious node $n'$ has the value of $\mathbf{Y}_i(x)$ at point $\alpha_{n'}$. Hence, malicious adversaries has the value of $\mathbf{Y}_i(x)$ at $t$ points. In addition we know that the value of $\mathbf{Y}_i(0)$ is equal to 0. By using Lagrange interpolation rules, malicious can construct $\mathbf{Y}_i(x)$ their self. Hence they can compute the value of $\mathbf{Y}(x)=[\mathbf{Y}_1(x),\mathbf{Y}_2(x),\dots,\mathbf{Y}_m(x)]$ at any point. Thus, they can not gain any additional information beyond what they have had before.

As shown, in this step, malicious workers can not gain any independent equation with equations, received in Step 1. Thus if we call the set of messages, which the malicious workers receive from other parties as $\mathcal{Y}_{\mathcal{S}}^{{(\delta,2)}}$, we have
\begin{align}
H(\mathcal{Y}_{\mathcal{S}}^{{(\delta,2)}}|\mathcal{X},  \mathcal{Y}_{\mathcal{S}}^{(1)},\mathcal{Y}_{\mathcal{S}}^{(2)},...,\mathcal{Y}_{\mathcal{S}}^{(\delta-1)},\mathcal{Y}_{\mathcal{S}}^{(\delta,\mathbf{A})},\mathcal{Y}_{\mathcal{S}}^{(\delta,\mathbf{B})})=0.
\end{align}

\subsubsection{Constructing the product of subshares:}

\textbf{Step 1}: First, each honest worker $n$ constructs $\mathbf{O}_l^{(n)}(x)= \mathbf{R}_{l,0}+\mathbf{R}_{l,1}x+\dots+\mathbf{R}_{l,t-1}x^{t-1}+\mathbf{D}^{(n)}_{l}x^t$ of degree $t$, where $0 \leq l \leq m+t-2$. Let us call the set of all random matrices which are used in this step as  $\mathcal{R}^{(\delta,O)}=\{\mathbf{R}_{l,j}|0 \leq l \leq m+t-2 \text{ and } 0 \leq j \leq t-1 \}$ and let us define $\mathcal{Y}_{\mathcal{S}}^{(\delta,O)}$ as the set of all messages which malicious workers receive from the other nodes. Malicious nodes have the value of $\mathbf{O}_l^{(n)}$ at $t$ points for each honest node $n$ and $0 \leq l  \leq m+t-2$. On the other hand, the number of random matrices used in $\mathbf{O}_l^{(n)}$ is $t$. Thus, the number of independent messages $|\mathcal{Y}_{\mathcal{S}}^{(\delta,O)}|$ is equal with the number of random unknown matrices $|\mathcal{R}^{(\delta,O)}|$. Due to the fact that $\mathcal{Y}_{\mathcal{S}}^{(\delta,O)}$ is a linear function of $\mathcal{R}^{(\delta,O)}$, Hence, there is one to one mapping between elements of $\mathcal{R}^{(\delta,O)}$ and $\mathcal{Y}_{\mathcal{S}}^{(\delta,O)}$. It means that,
\begin{align}
H(\mathcal{Y}_{\mathcal{S}}^{(\delta,O)}|\mathcal{X},  \mathcal{Y}_{\mathcal{S}}^{(1)},\mathcal{Y}_{\mathcal{S}}^{(2)},...,\mathcal{Y}_{\mathcal{S}}^{(\delta-1)},\mathcal{Y}_{\mathcal{S}}^{(\delta,\mathbf{A})},&\mathcal{Y}_{\mathcal{S}}^{(\delta,\mathbf{B})},\mathcal{Y}_{\mathcal{S}}^{(\delta,2)})\\&= H(\mathcal{R}^{(\delta,O)}|\mathcal{X},  \mathcal{Y}_{\mathcal{S}}^{(1)},\mathcal{Y}_{\mathcal{S}}^{(2)},...,\mathcal{Y}_{\mathcal{S}}^{(\delta-1)},\mathcal{Y}_{\mathcal{S}}^{(\delta,\mathbf{A})},\mathcal{Y}_{\mathcal{S}}^{(\delta,\mathbf{B})},\mathcal{Y}_{\mathcal{S}}^{(\delta,2)})\\
&= H(\mathcal{R}^{(\delta,O)})
 \overset{a}{\geq} H(\mathcal{Y}_{\mathcal{S}}^{(\delta,O)}),
\end{align}
where $(a)$ is valid because $\mathcal{Y}_{\mathcal{S}}^{(\delta,O)}$ and $\mathcal{R}^{(\delta,O)}$ have the same size, while $\mathcal{R}^{(\delta,O)}$ is selected uniformly and independently at random from the set of all matrices with the same size.

\textbf{Step 2}: In this step, each honest node $n$, invokes Algorithm \ref{ProtocolEVSS} to share $\mathbf{C}_n(x)$. We define $\mathcal{Y}_{\mathcal{S}}^{(\delta,C)}$ as the set of all messages which malicious nodes receive from the other nodes in this step. Similar to Step 2 of Subsection \ref{privacysubshare}, by partitioning random matrices into $\mathcal{R}^{(\delta,C)}$ and $\mathcal{R'}^{(\delta,C)}$, we can prove that,

\begin{align}
H(\mathcal{Y}_{\mathcal{S}}^{(\delta,C)}|\mathcal{X},  \mathcal{Y}_{\mathcal{S}}^{(1)},\mathcal{Y}_{\mathcal{S}}^{(2)},...,\mathcal{Y}_{\mathcal{S}}^{(\delta-1)},\mathcal{Y}_{\mathcal{S}}^{(\delta,\mathbf{A})},\mathcal{Y}_{\mathcal{S}}^{(\delta,\mathbf{B})},\mathcal{Y}_{\mathcal{S}}^{(\delta,2)},\mathcal{Y}_{\mathcal{S}}^{(\delta,O)}) \geq H(\mathcal{Y}_{\mathcal{S}}^{(\delta,C)}).
\end{align}

\textbf{Step 3}: This step is needed for verification of complaints in constructing the product of subshares phase. Assume that node $n'$ sends a complaint message. Then all nodes compute values of $\mathbf{A}_n(\alpha_{n'}),\mathbf{B}_n(\alpha_{n'}),$ $\mathbf{O}^{(n)}_0(\alpha_{n'}),\dots,$ $\mathbf{O}^{(n)}_{m+t-2}(\alpha_{n'})$ and $\mathbf{C}_n(\alpha_{n'})$, together by invoking Algorithm \ref{protocolEval}.  Sending complaint message from node $n'$ means that at least one node $n'$ or the dealer (node $n$) is malicious. So these values are known to the malicious nodes. Hence, these values cannot add any additional information to the malicious nodes. Thus, if we call $\mathcal{Y}_{\mathcal{S}}^{(\delta,E)}$ as the set of all messages which malicious parties receive in this step, we have

\begin{align}
H(\mathcal{Y}_{\mathcal{S}}^{(\delta,E)}|\mathcal{X},  \mathcal{Y}_{\mathcal{S}}^{(1)},\mathcal{Y}_{\mathcal{S}}^{(2)},...,\mathcal{Y}_{\mathcal{S}}^{(\delta-1)},\mathcal{Y}_{\mathcal{S}}^{(\delta,\mathbf{A})},\mathcal{Y}_{\mathcal{S}}^{(\delta,\mathbf{B})},\mathcal{Y}_{\mathcal{S}}^{(\delta,2)},\mathcal{Y}_{\mathcal{S}}^{(\delta,O)},\mathcal{Y}_{\mathcal{S}}^{(\delta,C)})=0.
\end{align}

It should be mentioned that, in the process of complaints verification in this step, each node $i$ subshares its share, i.e., $\mathbf{A}_n(\alpha_{i}),\mathbf{B}_n(\alpha_{i}),\mathbf{O}^{(n)}_0(\alpha_{i}),\dots,\mathbf{O}^{(n)}_{m+t-2}(\alpha_{i})$ and $\mathbf{C}_n(\alpha_{i})$.  We define $\mathcal{Y}_{\mathcal{S}}^{(\delta,E')}$ as the set of all messages which malicious nodes receive from other parties in this step. Similar to Step 2 of \ref{privacysubshare}, by partitioning random matrices into $\mathcal{R}^{(\delta,E')}$ and $\mathcal{R'}^{(\delta,E')}$, we can prove that,

\begin{align}
H(\mathcal{Y}_{\mathcal{S}}^{(\delta,E')}|\mathcal{X},  \mathcal{Y}_{\mathcal{S}}^{(1)},\mathcal{Y}_{\mathcal{S}}^{(2)},...,\mathcal{Y}_{\mathcal{S}}^{(\delta-1)},\mathcal{Y}_{\mathcal{S}}^{(\delta,\mathbf{A})},\mathcal{Y}_{\mathcal{S}}^{(\delta,\mathbf{B})},\mathcal{Y}_{\mathcal{S}}^{(\delta,2)},\mathcal{Y}_{\mathcal{S}}^{(\delta,O)},\mathcal{Y}_{\mathcal{S}}^{(\delta,C)},\mathcal{Y}_{\mathcal{S}}^{(\delta,E)})\geq H(\mathcal{Y}_{\mathcal{S}}^{(\delta,E')}).
\end{align}

\subsection{Proof of Privacy for Transposing:}
As it was mentioned in Subsection \ref{Sharesoftrransposedmatrix}, to be able to derive the shares of matrix $\Ab^T$ from shares of matrix $\Ab$, we need to follows a 2-step method as explained in \ref{Sharesoftrransposedmatrix}. In the first step, each node $n$ needs to compute $\Ab^{(0)}(x),\Ab^{(2)}(x),\dots,\Ab^{(m-1)}(x)$ at point $\alpha_n$ for each $n$. To do this, all nodes use subshare of shares algorithm (Algorithm \ref{protocolsubshare}). The privacy proof of subshare of shares algorithm is similar to privacy of subshare of shares which is shown in Subsection \ref{privacysubshare}. In the second step, there is no communication among nodes. Thus there is no data leakage at this step.

\subsection{Proof of Privacy for Changing the Parameter of the Shares:}
Similar to the previous subsection, all nodes need to follow a 2-step algorithm. First, nodes need to use subshare of shares algorithm (Algorithm \ref{protocolsubshare}). The privacy of subshare of shares algorithm is proved in Subsection \ref{privacysubshare}. And in the second step, there is no communication among nodes. Thus there is no data leakage at this step.

\subsection{Proof of Privacy for One Round in  the Algorithm:}
To be able to prove that the mutual information of input matrices and the set of all messages that malicious nodes receive in different rounds is 0, we need to prove that for each round $\delta$, 
\begin{align}
\label{mutual}
H(\mathcal{Y}_{\mathcal{S}}^{(\delta)}|\mathcal{X},  \mathcal{Y}_{\mathcal{S}}^{(1)},\mathcal{Y}_{\mathcal{S}}^{(2)},...,\mathcal{Y}_{\mathcal{S}}^{(\delta-1)})=H(\mathcal{Y}_{\mathcal{S}}^{(\delta)}).
\end{align}

Here we prove \eqref{mutual} for the cases, where round $\delta$ is multiplication of two matrices. For other operations, we can follow the same steps.

\begin{align}
H(\mathcal{Y}_{\mathcal{S}}^{(\delta)}|\mathcal{X},  \mathcal{Y}_{\mathcal{S}}^{(1)},\mathcal{Y}_{\mathcal{S}}^{(2)},...,\mathcal{Y}_{\mathcal{S}}^{(\delta-1)})&=
H(\mathcal{Y}_{\mathcal{S}}^{(\delta,\mathbf{A})},\mathcal{Y}_{\mathcal{S}}^{(\delta,\mathbf{B})},\mathcal{Y}_{\mathcal{S}}^{(\delta,O)},\mathcal{Y}_{\mathcal{S}}^{(\delta,C)},\mathcal{Y}_{\mathcal{S}}^{(\delta,E')}|\mathcal{X},  \mathcal{Y}_{\mathcal{S}}^{(1)},\mathcal{Y}_{\mathcal{S}}^{(2)},...,\mathcal{Y}_{\mathcal{S}}^{(\delta-1)})\\&=
H(\mathcal{Y}_{\mathcal{S}}^{(\delta,\mathbf{A})}|\mathcal{X},  \mathcal{Y}_{\mathcal{S}}^{(1)},\mathcal{Y}_{\mathcal{S}}^{(2)},...,\mathcal{Y}_{\mathcal{S}}^{(\delta-1)})\\&+
H(\mathcal{Y}_{\mathcal{S}}^{(\delta,\mathbf{B})}|\mathcal{X},  \mathcal{Y}_{\mathcal{S}}^{(1)},\mathcal{Y}_{\mathcal{S}}^{(2)},...,\mathcal{Y}_{\mathcal{S}}^{(\delta-1)},\mathcal{Y}_{\mathcal{S}}^{(\delta,\mathbf{A})})\\&+
H(\mathcal{Y}_{\mathcal{S}}^{(\delta,O)}|\mathcal{X},  \mathcal{Y}_{\mathcal{S}}^{(1)},\mathcal{Y}_{\mathcal{S}}^{(2)},...,\mathcal{Y}_{\mathcal{S}}^{(\delta-1)},\mathcal{Y}_{\mathcal{S}}^{(\delta,\mathbf{A})},\mathcal{Y}_{\mathcal{S}}^{(\delta,\mathbf{B})})\\&+
H(\mathcal{Y}_{\mathcal{S}}^{(\delta,C)}|\mathcal{X},  \mathcal{Y}_{\mathcal{S}}^{(1)},\mathcal{Y}_{\mathcal{S}}^{(2)},...,\mathcal{Y}_{\mathcal{S}}^{(\delta-1)},\mathcal{Y}_{\mathcal{S}}^{(\delta,\mathbf{A})},\mathcal{Y}_{\mathcal{S}}^{(\delta,\mathbf{B})},\mathcal{Y}_{\mathcal{S}}^{(\delta,O)})\\&+
H(\mathcal{Y}_{\mathcal{S}}^{(\delta,E')}|\mathcal{X},  \mathcal{Y}_{\mathcal{S}}^{(1)},\mathcal{Y}_{\mathcal{S}}^{(2)},...,\mathcal{Y}_{\mathcal{S}}^{(\delta-1)},\mathcal{Y}_{\mathcal{S}}^{(\delta,\mathbf{A})},\mathcal{Y}_{\mathcal{S}}^{(\delta,\mathbf{B})},\mathcal{Y}_{\mathcal{S}}^{(\delta,O)},\mathcal{Y}_{\mathcal{S}}^{(\delta,C)})\\ & \geq
H(\mathcal{Y}_{\mathcal{S}}^{(\delta,A)})+H(\mathcal{Y}_{\mathcal{S}}^{(\delta,B)})+H(\mathcal{Y}_{\mathcal{S}}^{(\delta,O)})+H(\mathcal{Y}_{\mathcal{S}}^{(\delta,C)})+H(\mathcal{Y}_{\mathcal{S}}^{(\delta,E')}) \\ &\geq H(\mathcal{Y}_{\mathcal{S}}^{(\delta,A)},\mathcal{Y}_{\mathcal{S}}^{(\delta,B)},\mathcal{Y}_{\mathcal{S}}^{(\delta,O)},\mathcal{Y}_{\mathcal{S}}^{(\delta,C)},\mathcal{Y}_{\mathcal{S}}^{(\delta,E')})
\end{align}

Thus, we have
\begin{align*}
H(\mathcal{Y}_{\mathcal{S}}^{(\delta)}|\mathcal{X},  \mathcal{Y}_{\mathcal{S}}^{(1)},\mathcal{Y}_{\mathcal{S}}^{(2)},...,\mathcal{Y}_{\mathcal{S}}^{(\delta-1)})=H(\mathcal{Y}_{\mathcal{S}}^{(\delta)}).
\end{align*}

\subsection{Proof of Privacy for the Algorithm:}
Let us define $\mathcal{Y}_{\mathcal{S}}^{(\delta)}$ as the set of all messages that malicious workers achieve in round $\delta$. 
Assume that the algorithm is executed in $\Delta$ rounds. To prove that the algorithm has no data leakage, it is sufficient to prove that $I(  \mathcal{Y}_{\mathcal{S}}^{1},\mathcal{Y}_{\mathcal{S}}^{2},...,\mathcal{Y}_{\mathcal{S}}^{\Delta};\mathcal{X})=0$. One can see that
\begin{align*}
H(  \mathcal{Y}_{\mathcal{S}}^{(1)},\mathcal{Y}_{\mathcal{S}}^{(2)},...,\mathcal{Y}_{\mathcal{S}}^{(\Delta)}|\mathcal{X})
=&
\displaystyle\sum_{\delta=1}^{\Delta}H(\mathcal{Y}_{\mathcal{S}}^{(\delta)}| \mathcal{X}, \mathcal{Y}_{\mathcal{S}}^{(1)},\mathcal{Y}_{\mathcal{S}}^{(2)},...,\mathcal{Y}_{\mathcal{S}}^{(\delta-1)}) \\
\geq &  \displaystyle\sum_{\delta=1}^{\Delta}H(\mathcal{Y}_{\mathcal{S}}^{(\delta)})
\geq H(  \mathcal{Y}_{\mathcal{S}}^{(1)},\mathcal{Y}_{\mathcal{S}}^{(2)},...,\mathcal{Y}_{\mathcal{S}}^{(\Delta)}).
\end{align*}
Hence, obviously
\begin{align*}
I(  \mathcal{Y}_{\mathcal{S}}^{(1)},\mathcal{Y}_{\mathcal{S}}^{(2)},...,\mathcal{Y}_{\mathcal{S}}^{(\Delta)};\mathcal{X})=H(  \mathcal{Y}_{\mathcal{S}}^{(1)},\mathcal{Y}_{\mathcal{S}}^{(2)},...,\mathcal{Y}_{\mathcal{S}}^{(\Delta)})-H(  \mathcal{Y}_{\mathcal{S}}^{(1)},\mathcal{Y}_{\mathcal{S}}^{(2)},...,\mathcal{Y}_{\mathcal{S}}^{(\Delta)}|\mathcal{X})=0.
\end{align*}

\bibliographystyle{ieeetr}

\end{document}